\documentclass[sigconf, 9pt]{acmart}

\usepackage[english]{babel}
\usepackage[utf8]{inputenc}
\usepackage{url}
\usepackage{xcolor}
\usepackage{mathtools} 
\usepackage{siunitx} 
\usepackage{xspace}
\usepackage{svg}

\usepackage{tikz}

\definecolor{circlebg}{RGB}{245,245,245}
\definecolor{mutedblack}{RGB}{51,51,51}
\newcommand*\circled[1]{\tikz[baseline=(char.base)]{
    \node[shape=circle, draw=mutedblack, text=mutedblack, fill=circlebg, inner sep=0.5pt] (char) {#1};}%
}

\newcommand{\ie}{{\em i.e.,\/ }}
\newcommand{\eg}{{\em e.g.,\/ }}
\newcommand{\vs}{{\em vs.\/ }}

\newcommand{\pb}[1]{\vspace{0.75ex}\noindent{\bf \em #1}\hspace*{.3em}}

\newcommand{\one}{({\em i}\/)\xspace}
\newcommand{\two}{({\em ii}\/)\xspace}
\newcommand{\three}{({\em iii}\/)\xspace}
\newcommand{\four}{({\em iv}\/)\xspace}

\newcommand{\six}

\setcopyright{acmcopyright}

\title[Design and Evaluation of IPFS]{Design and Evaluation of IPFS:\\A Storage Layer for the Decentralized Web}

\author{Dennis Trautwein}
\orcid{0000-0002-8567-2353}

\affiliation{%
  \institution{Protocol Labs \&}
}
\affiliation{%
   \institution{University of Göttingen}
}
\email{dennis.trautwein@protocol.ai}

\author{Aravindh Raman}
\affiliation{%
  \institution{Telefonica Research}
}
\email{aravindh.raman@telefonica.com}

\author{Gareth Tyson}
\affiliation{%
  \institution{Hong Kong University of Science \& Technology (GZ)}
}
\email{gtyson@ust.hk}

\author{Ignacio Castro}
\affiliation{%
  \institution{Queen Mary University of London}
}
\email{i.castro@qmul.ac.uk}

\author{Will Scott}
\affiliation{%
  \institution{Protocol Labs}
}
\email{will@protocol.ai}

\author{Moritz Schubotz}
\orcid{0000-0001-7141-4997}
\affiliation{%
  \institution{FIZ Karlsruhe -- Leibniz Institute for Information Infrastructure}
}
\email{moritz.schubotz@fiz-karlsruhe.de}

\author{Bela Gipp}
\orcid{0000-0001-6522-3019}
\affiliation{%
  \institution{University of Göttingen}
}
\email{gipp@uni-goettingen.de}

\author{Yiannis Psaras}
\affiliation{%
  \institution{Protocol Labs}
}
\email{yiannis@protocol.ai}



\renewcommand{\shortauthors}{Trautwein et al.}

\begin{document}

\acmYear{2022}\copyrightyear{2022}
\setcopyright{rightsretained}
\acmConference[SIGCOMM '22]{ACM SIGCOMM 2022 Conference}{August 22--26, 2022}{Amsterdam, Netherlands}
\acmBooktitle{ACM SIGCOMM 2022 Conference (SIGCOMM '22), August 22--26, 2022, Amsterdam, Netherlands}
\acmPrice{}
\acmDOI{10.1145/3544216.3544232}
\acmISBN{978-1-4503-9420-8/22/08}

\begin{abstract}

Recent years have witnessed growing consolidation of web operations. For example, the majority of web traffic now originates from a few organizations, and even micro-websites often choose to host on large pre-existing cloud infrastructures.
In response to this, the ``Decentralized Web'' attempts to distribute ownership and operation of web services more evenly. 
This paper describes the design and implementation of the largest and most widely used Decentralized Web platform --- the InterPlanetary File System (IPFS) --- an open-source, content-addressable peer-to-peer network that provides distributed data storage and delivery. IPFS has millions of daily content retrievals and already underpins dozens of third-party applications. 
This paper evaluates the performance of IPFS by introducing a set of measurement methodologies that allow us to uncover the characteristics of peers in the IPFS network. We reveal presence in more than 2700 Autonomous Systems and 152 countries, the majority of which operate outside large central cloud providers like Amazon or Azure. We further evaluate IPFS performance, showing that both publication and retrieval delays are acceptable for a wide range of use cases. 
Finally, we share our datasets, experiences and lessons learned.

\end{abstract}

\maketitle

\section{Introduction}
\label{sec:introduction}

Economies of scale and technical innovations, such as cloud computing, have led to a growing centralization of web systems~\cite{bommelaer2019global}. For example, recent trends in name resolution, content hosting, routing~\cite{castro2013using,bottger2018elusive}, protocol development~\cite{holz2020tracking, mcquistin2021characterising, khare2022web} and certificate authorities all point towards the consolidation of ownership and operation~\cite{raman2019challenges}.
An administrator establishing a new website will likely co-locate their server on a cloud platform such as Amazon EC2; utilize a third-party DNS provider such as GoDaddy; serve their content via a Content Delivery Network like Akamai;
and rely on certificates issued by Let's Encrypt. Although each of these services is well-engineered and highly performant, they nevertheless represent single points of organizational failure. In the most extreme cases, such players have gained near-monopoly status and triggered widespread chaos during outages (\eg OVHcloud, Cloudflare, AWS)~\cite{ovhOutage, cloudFlareDNS, awsOutage}. The monetary costs incurred during outages are enormous, with Amazon's eCommerce platform reportedly losing over \SI{66000}[\$]{} per minute during an outage in 2013~\cite{amazon_loss}.

In response to this, there has been a growing movement, colloquially referred to as the ``\emph{Decentralized Web}''. This encompasses an array of technologies that strive to provide greater control for users.
These technologies tend to rely on open-source, community-led software implementations that decentralize traditional web functionality (\eg name lookup, hosting, certification), such that no individual administrative entity could hamper overall operations or design decisions. 
A number of successful projects have already deployed decentralized systems that offer commonly used services, \eg Mastodon for micro-blogging or PeerTube for video sharing. However, at the core of any web platform is storing and serving media objects at scale. We argue that by decentralizing these core functions, many other applications could readily be built atop without needing to handle the complexity of decentralization themselves. 

The \emph{InterPlanetary File System (IPFS)} project aims to achieve this:
it is an entirely decentralized content-addressable media object storage and retrieval platform.
IPFS is a community-driven, open source effort, which is vital for ensuring community buy-in and creating an open platform for design innovation. IPFS covers 176 git repositories, across which there have been 60.4\,k commits by 1185 code contributors, covering 400+ organizations including universities, start-ups and large corporations. This paper reports on our experiences in \emph{Protocol Labs}, driving forward the IPFS effort. Protocol Labs is the largest supporter of the IPFS project, employing or funding most of the full-time contributors. Protocol Labs is also the largest contributor to the open-source codebase, covering \SI{62}{\percent} of git commits and \SI{75.4}{\percent} lines of code. It is worth noting, however, that large codebase decisions and roadmap setting is led through a public voting process.
Thus, we emphasize that the design and implementation work reported in this paper stems from countless community contributions.

IPFS is seeing widespread uptake with more than 3\,M web client accesses and beyond 300\,k unique nodes serving content in the peer-to-peer (P2P) network every week.
IPFS currently underpins various other Decentralized Web applications, including social networking and discussion platforms (Discussify, Matters News), data storage solutions (Space, Peergos, Temporal), content search (Almonit, Deece), messaging (Berty), content streaming (Audius, Watchit), and e-commerce (Ethlance, dClimate)~\cite{ipfsecosystem}.
Support for accessing IPFS has further been integrated into mainstream browsers such as Opera and Brave, allowing widespread and easy uptake. 

In this paper, we present the design and implementation of IPFS. 
At its core, IPFS relies on four main concepts: 
\one~\emph{Content-based addressing}: unlike HTTP, IPFS detaches object names from host location --- enabling objects to be served from any peer;
\two~\emph{Decentralized object indexing}: IPFS relies on a decentralized P2P overlay for indexing all available locations from which objects can be retrieved reducing the impact of technical or organizational failure;
\three~\emph{Immutability and self-certification:} IPFS relies on cryptographic hashing to self-certify objects, removing the need for certificate-based authentication, hence, providing verifiability;
and
\four~\emph{Open participation:} anybody can deploy an IPFS node and participate in the network without requiring special permissions or privileges.

The contributions of this paper are as follows:

\begin{enumerate}
    \item We present the design and implementation of IPFS (Section \ref{sec:ipfs_fundamentals}), detailing how it publishes (Section \ref{sec:content-publication}) and retrieves (Section \ref{sec:content-routing}) content at scale. 

    \item We propose three complementary measurement methodologies that provide vantage into the deployment, usage and performance of the IPFS network (Section \ref{sec:evaluation_data}). This is vital due to the decentralized nature of IPFS. As no individual entity operates the entirety of IPFS, we use these techniques to quantify IPFS across a number of dimensions. We make our datasets and tooling publicly available.
    
    \item We utilize the above methodologies to evaluate the deployment success of IPFS (Section \ref{sec:deployment}). We find that IPFS infrastructure has been deployed in over \num{2700} Autonomous Systems, across 464\,k IP addresses. This covers 152 countries, with the majority hosted in the US and China. We further observe widespread usage by clients with 7.1 million content retrievals seen from a single vantage point on one day alone.
    
    \item We finally present a performance evaluation of IPFS (Section \ref{sec:evaluation}). We show that, although content retrievals in IPFS are slower than direct HTTP access, delays are still reasonable for a number of use cases. For example, $3/4$ of retrievals from Europe are under 2 seconds. This includes looking up the content host and fetching a 0.5 MB file. To improve performance, we show how the introduction of gateway caching can substantially reduce retrieval latency with \SI{76}{\percent} of requests being served in under \SI{250}{\milli\second}.
    
\end{enumerate}

\section{IPFS Fundamentals}
\label{sec:ipfs_fundamentals}

We start by providing an overview of the core building blocks of IPFS. Namely, how IPFS \one~addresses content; \two~addresses peers; and \three~indexes content, to enable distributed lookups that map content identifiers to peers hosting the object.

\subsection{Content Addressing}
\label{sec:ipfs_fundamentals_content_addressing}

At the core of IPFS is a content-based addressing scheme using unique hash-based \textbf{Content Identifiers (CIDs)}, similar to BitTorrent~\cite{bittorrent} or Content-Centric Networking~\cite{10.1145/2656877.2656887} (see related work in Section~\ref{sec:related-work}).
CIDs are the base primitive that decouple a name for content from the storage location. 
In contrast, location-based systems, such as HTTP, bind content addresses (URLs) to their primary host. This fundamental design decision enables the decentralization of content storage, content delivery, and address management. In addition, by decoupling the content address from its storage location, CIDs prevent vendor lock-in and remove the need for central authorities to handle address allocation.

\begin{figure}[t]
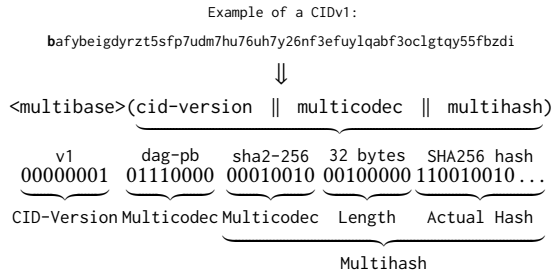

    \centering
    \texttt{\scriptsize Example of a CIDv1:}\\
    \texttt{\scriptsize \textbf{b}afybeigdyrzt5sfp7udm7hu76uh7y26nf3efuylqabf3oclgtqy55fbzdi}\\[.2em]
    $\Downarrow$\\[.2em]
    \texttt{\small <multibase>($\underbrace{\texttt{cid-version}~\parallel~\texttt{multicodec}~\parallel~\texttt{multihash}}$)}\\[.2em]
    $
    \underbrace{\overset{\texttt{v1}}{00000001}}_{\texttt{CID-Version}}
    \underbrace{\overset{\texttt{dag-pb}}{01110000}}_{\texttt{Multicodec}}
    \underbrace{\underbrace{\overset{\texttt{sha2-256}}{00010010}}_{\texttt{Multicodec}}
    \underbrace{\overset{\text{\texttt{32 bytes}}}{00100000}}_{\texttt{Length}}\underbrace{\overset{\texttt{SHA256 hash}}{110010010\dots}}_{\texttt{Actual Hash}}}_{\texttt{Multihash}}
    $
    \vspace{-0.2cm}
    \caption{Structure of a CID.}
    \label{fig:cid_structure}
    \vspace{-0.5cm}
\end{figure}

Figure~\ref{fig:cid_structure} shows an example of a CID and its structure. It relies on a set of self-describing data representation protocols~\cite{multiformats} and is composed of the following four fields:

\begin{description}
    \item[Multibase prefix:]
        Indicates one of the 24 currently supported base-encodings with which the binary CID has been encoded (``b" for base32 in Figure~\ref{fig:cid_structure}).
    
    \item[CID-Version identifier:]
        Indicates the CID version (v1). Currently, two versions exist (v0 and v1).
    
    \item[Multicodec identifier:]
        Specifies how the addressed data has been encoded (protobuf, json, cbor, etc.).
    
    \item[Multihash:]
        A self-describing hash-digest of the addressed data. 
        The Multihash includes metadata indicating the hash function used (default sha2-256) and the length (default 32 bytes) of the actual content hash. The term \textbf{Multi}hash stems from the fact that it can support any hashing algorithm.
\end{description}
When content is added to IPFS, it is split into chunks (default 256\,kB) each of which is assigned its own CID.
The CID of each chunk results from hashing its content and adding the above metadata.
Once all chunks have a CID, IPFS constructs a \textbf{Merkle Directed Acyclic Graph (DAG)} of the file~\cite{merkleDAG}. This Merkle DAG is the form in which the file is provided by the original content publisher.
A Merkle-DAG is a data structure similar to a Merkle-tree but without balance requirements.
The root node combines all CIDs of its descendant nodes and forms the final content CID (commonly called \textit{root CID}).
In Merkle-DAGs, a node is allowed to have multiple parents, an important property that allows for chunk de-duplication.
In turn, content de-duplication means that the same content does not need to be stored or transmitted twice, saving both storage and bandwidth resources. Further, Merkle DAGs are agnostic to where the content is stored. Thus, they do not need to be updated when a file is replicated on or deleted from nodes in the network.

Thanks to their hash-based structure, CIDs are
immutable and self-certifying, 
\ie content cannot be altered without modifying its CID. 
This enables self-verification by comparing the CID with the hash of the content itself.
Clearly, this property becomes a challenge for dynamically changing digital objects, which we address in Section~\ref{sec:content-mutability}.

\subsection{Peer Addressing}
%
Upon joining the IPFS network by connecting to a set of canonical bootstrap peers, peers generate a public-private key pair.
Every peer in the IPFS network is identified by its unique \textbf{PeerID}, which is the hash of its public key (represented as a \textit{Multihash}). The PeerID remains the same, unless the node operator chooses to change it manually.
When establishing a secure communication channel, the PeerID is used to verify that the public key used to secure the channel is the same as the one used to identify the peer.

In order to represent the locations of remote peers, IPFS relies on \textbf{Multiaddresses}.
A Multiaddress is a self-describing, human-readable, hierarchically-separated sequence of protocol choices.
The term \textbf{Multi}address stems from the fact that the format allows multiple protocols and address types to be included.
Each Multiaddress describes an endpoint enabling a peer to be interacted with. IPFS encompasses multiple protocols, from the network layer up to the application layer.

\begin{figure}
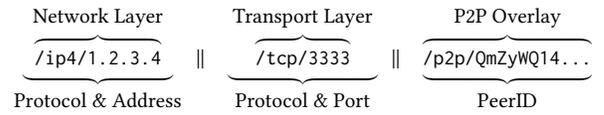

    \centering
    \small
    $
                    \underbrace{\overbrace{\hbox{\texttt{/ip4/1.2.3.4}}}^{\parbox{2cm}{\centering Network Layer}}}_{\parbox{2.4cm}{\centering Protocol \& Address}} 
    \parallel \ \ \ \underbrace{\overbrace{\hbox{\texttt{/tcp/3333}}}^{\parbox{2cm}{\centering Transport Layer}}}_{\hbox{Protocol \& Port}}\
    \parallel \ \ \ \underbrace{\overbrace{\hbox{\texttt{/p2p/QmZyWQ14...}}}^{\parbox{2cm}{\centering P2P Overlay}}}_{\hbox{PeerID}}\ 
    $
    \caption{Structure of a Multiaddress.}
    \label{fig:structure_multiaddress}
\end{figure}

Figure~\ref{fig:structure_multiaddress} presents the structure of a Multiaddress, showing the network and transport protocols for the communication (\texttt{IPv4} and \texttt{TCP}) their corresponding location-based address information (IP address \texttt{1.2.3.4} and TCP port number \texttt{3333}) followed by the protocol to address one particular peer (\texttt{p2p}) and its PeerID (\texttt{QmZyWQ14...}). As a result, Multiaddresses point to remote processes by encoding multiple layers of addressing information into a path representation. A Multiaddress uses this construct for two reasons. First, not all IPFS nodes share the same subset of protocols. Multiaddresses allow nodes to know if they will be able to connect to a remote peer before attempting the connection. Second, the extensible syntax of Multiaddresses allows for intermediate relaying of communication through prefixing peer addresses. This is used to proxy messages to in-browser nodes that cannot be directly contacted.

\subsection{Content Indexing}
\label{sec:dht}
%
%
%
To publish or retrieve an object, it is necessary to create a mapping between a CID and a PeerID that can provide the object (including its Multiaddress). In order to operate in a decentralized fashion and support content and peer discovery, these mappings are indexed on a Distributed Hash Table (DHT), which exposes simple \texttt{PUT} and \texttt{GET} primitives. IPFS's DHT is based on Kademlia~\cite{maymounkov2002kademlia}, similar to that used by the BitTorrent Mainline DHT~\cite{gnutella, bittorrent}. 
CIDs and PeerIDs reside in a common 256-bit key space by using the SHA256 hashes of their binary representations (see Figure~\ref{fig:cid_structure}) as indexing keys.

Based on our practical experiences with the live network, we have made a number of tweaks compared to the original Kademlia specification.
Nodes in the DHT use 256-bit SHA256 keys instead of the 160-bit SHA1 keys.
This is to anticipate advances in deliberate hash collisions~\cite{Wang2005}.
We also maintain $i=256$ buckets of $k-$nodes each (where $k=20$) to split the hash space.
Finally, we employ reliable transport protocols such as TCP and QUIC (instead of UDP)~\cite{maymounkov2002kademlia}, as this makes connection management in the implementation more straightforward.



\makeatletter
    \DeclareRobustCommand*{\escapeus}[1]{%
        \begingroup\@activeus\scantokens{#1 }\endgroup}
    \begingroup\lccode`\~=`\_\relax
        \lowercase{\endgroup\def\@activeus{\catcode`\_=\active \let~\_}}
\makeatother

New peers join the DHT as either \textit{DHT Servers}, if they have public IP connectivity, or as \textit{DHT Clients}, if they are not publicly reachable, \eg because they are behind a Network Address Translation (NAT) device.
IPFS differentiates between DHT clients and servers through a simple technique called \textit{Autonat}~\cite{autonat}. Autonat works as follows: new peers join by default as clients and immediately ask other peers in the network to initiate connections back to them. If more than three peers can connect to the newly joining peer, then the new peer upgrades its participation to act as a server node. If more than three peers cannot connect, the peer continues as a client. DHT Servers perform all network operations, \ie storing content, storing mapping records, and providing these to requesting peers. In contrast, DHT Clients only request records or content from the network but do not store or provide any of them. The DHT client/server distinction prevents unreachable peers from becoming part of other peers' routing tables, thus speeding up the publication and retrieval processes.
\section{IPFS in action}
\label{sec:ipfs_in_action}

\begin{figure}[t]
    \centering
    \includegraphics[width=\linewidth]{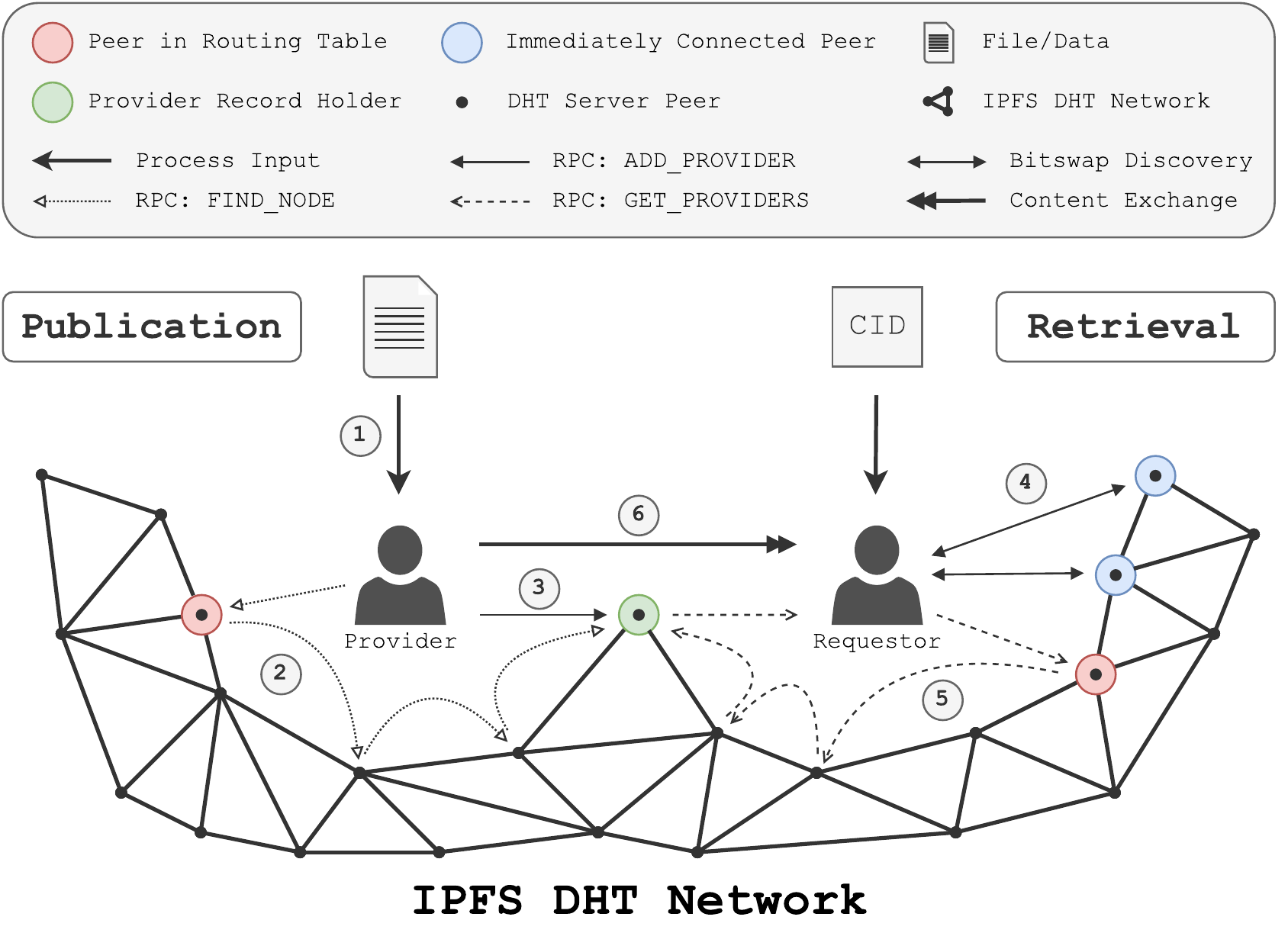}
    \caption{IPFS publication and retrieval. Publication involves:
    {\Large \textcircled{\Small 1}} Import content to \textit{Providers} local IPFS process and allocate CID {\Large \textcircled{\Small 2}} DHT Walk to find closest peers to CID {\Large \textcircled{\Small 3}} Store provider record with closest peers. Retrieval involves: {\Large \textcircled{\Small 4}} Opportunistic Bitswap requests to already connected peers for CID {\Large \textcircled{\Small 5}} DHT Walk to find a provider record storing peer {\Large \textcircled{\Small 6}} \textit{Requestor} connects to \textit{Provider} and fetches the content. The diagram omits \textit{Requestors} second DHT Walk to resolve \textit{Providers} PeerID to their network address.}
    
    \label{fig:ipfs_publication_retrieval}
\end{figure}

In this section we explain how content is published  (Section~\ref{sec:content-publication}), and how other peers are able to find and retrieve it (Section~\ref{sec:content-routing}). Both processes are depicted in Figure~\ref{fig:ipfs_publication_retrieval} and marked throughout this section. Then in Section~\ref{sec:content-mutability}, we describe how IPFS deals with mutable content.

\subsection{Content Publication}
\label{sec:content-publication}

To make content available in the IPFS network, the content is first imported to IPFS and allocated a CID \circled{1} (see Section~\ref{sec:ipfs_fundamentals_content_addressing}).

After content has been imported into the local IPFS instance, it is neither replicated nor uploaded to any external server.
To publish it, the host generates a \emph{provider record} and pushes it into the DHT.
This record maps the CID to its own PeerID. 
The provider record is stored on the $k=20$ closest peers in terms of their PeerIDs' XOR distance~\cite{maymounkov2002kademlia} from the SHA256 hash of the CID.
The record is replicated on $k$ peers to ensure that records remain available even if some of those $k$ peers leave the network.
20 is selected based on our practical experiences (as well as a recommendation in the original Kademlia specification~\cite{maymounkov2002kademlia}), serving as a compromise between excessive replication overhead and risking record deletion because of peer churn.
We explain how the 20 closest peers are discovered in Section~\ref{sec:content-routing}.

Once IPFS has found the closest peers \circled{2}, it attempts to store the provider record with them. It does so by establishing a network connection and then initiating an RPC \circled{3}. The process does not wait for a response from each peer but will instead perform the RPCs in a ``fire and forget'' fashion which will become relevant in the performance evaluation (Section~\ref{sec:evaluation}).
Note, any peer that later retrieves the data becomes a temporary (or permanent, if they so choose) content provider themselves by publishing a provider record pointing to their own node to the DHT. Peers retrieving the content do not need to trust the new providing peer but only verify that the data they were served matches the requested CID.

A peer must also publish its \emph{peer record}, which maps its own PeerID to any Multiaddresses associated with it. This is used by requesting peers to discover the underlying network address of the peer.
Publication of the peer record follows the same CID-to-PeerID procedure as described above and happens independently of any content publication.




Apart from the replication factor ($k=20$), provider records are associated with two other parameters: \one the \textit{republish interval}, by default set to \SI{12}{\hour}, to make sure that even if the original 20 peers previously responsible for keeping the provider record go offline, the provider will assign new ones within \SI{12}{\hour}; 
and 
\two the \textit{expiry interval}, by default set to \SI{24}{\hour}, to make sure that the publisher has not gone offline and is still willing to serve the content.
These settings aim to prevent the system from storing and providing stale records. 
It is worth noting that peers behind NATs cannot host content themselves.
Thus, third party hosts, commonly called \emph{pinning services} are used to publish content on behalf of NAT'ed end-users (usually for a fee). Although a NAT hole-punching solution is currently being developed~\cite{dcutr}, it is still under-test.


\subsection{Content Retrieval}
\label{sec:content-routing}
Once the provider and peer records have been published, users can retrieve the content. 
To retrieve the content, the requesting peer performs four steps: \one \textit{Content discovery:} identify PeerID(s) that host the content/CID;
\two \textit{Peer discovery:} map the PeerID to a Multiaddress, \eg an IP address;
\three \textit{Peer routing}: connect to the peer;
and 
\four~\textit{Content exchange:} fetch the content.

\pb{Content Discovery.}
Content discovery in IPFS is done primarily using the DHT (see Section~\ref{sec:ipfs_fundamentals}). 
However, before entering the DHT lookup, the requesting peer asks all peers it is already connected to for the desired CID \circled{4}. This is done (using the Bitswap protocol, discussed later in this section) in an opportunistic fashion.
It allows a node to resolve content faster in case a peer's immediate neighbours store the desired content. If that initial attempt is not successful, content discovery falls back to the DHT with a timeout of 1 second. 

The DHT implements \textit{multi-round iterative lookups} in order to resolve a CID to a peer's Multiaddresses, a process we refer to as a \emph{DHT walk} \circled{5}. When peer A issues a request for CID $x$, the request is forwarded to $\alpha=3$ nodes whose PeerIDs are closest to $x$ in peer A's routing table (as per the original Kademlia specification \cite{maymounkov2002kademlia}). The peers receiving those requests reply with the requested content if they have it. If they do not have the requested content, they reply with either the provider record that points to the PeerID that holds the requested item together with the peer's Multiaddress (if they have it), or with the peers it knows of whose PeerID is closer to $x$. The process continues until the node is returned with the PeerID that has previously declared to hold a copy of the requested CID through a published provider record.

\pb{Peer Discovery.}
After the content discovery phase, a client knows the PeerID(s) hosting the desired content. As mentioned earlier, PeerIDs need to be mapped to a physical network address by looking up the peer record.
This procedure is called \emph{Peer Discovery} and is carried out by  querying the DHT for a second time.


To further streamline the process, each IPFS node maintains an address book of up to 900 recently seen peers. Nodes check whether they already have an address for the PeerID they have discovered before performing any further lookups.


\pb{Peer Routing.}
Once the PeerID is resolved to a peer record, the requesting node will possess the Multiaddress(es) of the peer that appears to have the content. It therefore uses the list of addresses to connect to the desired peer.

\pb{Content Exchange.}
As provider peers are identified, content fetching is done using \textit{Bitswap}, a simple chunk exchange protocol \circled{6}. Bitswap issues requests for the content items in \textit{wantlists}. Requests are sent using an \texttt{IWANT-HAVE} message.
Recipient peers that have the block reply with a corresponding \texttt{IHAVE} message. The requesting peer finally responds with an \texttt{IWANT-BLOCK} message. Receipt of the requested block terminates the exchange.
Recall, the Bitswap protocol is also used to \emph{discover} content available on nearby neighbours opportunistically.
The full details of the Bitswap protocol design as well as a number of proposed optimizations can be found in~\cite{bitswap, RochaDP21}.

\subsection{Content Mutability}
\label{sec:content-mutability}

Given their hash-based structure, CIDs are permanent, immutable, and self-certifying.
This makes them ideal for decentralizing namespace management, as it avoids the need for a central coordinator. Furthermore, their immutability and self-certification make universal caching (\textit{i.e.,\/ } from any peer) possible.
However, this becomes problematic for handling dynamic content, \eg evolving text documents.
In order to cope with mutable content, IPFS provides the option of publishing content based on the hash of the publisher's public key (\ie PeerID) instead of the hash of the content (\ie CID) itself. Those, so called \textit{InterPlanetary Name System (IPNS) records}, map the CID of the publisher's public key to another CID signed by the corresponding private key. This way, content can be updated and obtain a different CID, but an immutable reference (\textit{i.e.,\/ } the CID of the publisher's public key) is created and used. As this mechanism makes use of additional constructs, we leave further details out and point the reader to \cite{ipns, dnslink} for more details.

\subsection{IPFS Gateways}
\label{sec:gateways}

To broaden access to IPFS-hosted content for users who have not installed IPFS software, the IPFS system is complemented with a \emph{gateway} model. Gateways offer (HTTP) entry points into IPFS, enabling users who do not run any IPFS software to access content. Our gateway implementation acts as a bridge: on one side is a DHT Server node, and on the other side is an nginx HTTP web server, which can receive \texttt{GET} requests containing the CID as the URL path, \ie \texttt{https://ipfs-gateway.io/ipfs/\{CID\}}. 

By embedding caching within the gateways, we further streamline performance by aggregating user demand. 
Each gateway server runs two forms of content storage: \one~the default nginx web cache, with a Least Recently Used replacement strategy; and \two~The IPFS node store, which holds content manually uploaded by the Web3 and NFT Storage Initiatives.\footnote{\url{https://web3.storage/}, \url{https://nft.storage}}
These allow third parties to pin content in the IPFS store of the gateway to make it persistently available. We emphasize that the gateways are entirely optional for the operation of the overall storage and retrieval network and are only used for content retrievals through the browser.

Protocol Labs operates two major gateways, and in total, there are 107 known gateways.\footnote{\url{https://ipfs.github.io/public-gateway-checker/}} 
Note, operating a gateway does not require authorization or permission by any entity, but in order for it to be useful for the network, it needs to be set up with a public IP address.

\begin{figure*}[ht]
    \centering
    \includegraphics[width=\textwidth]{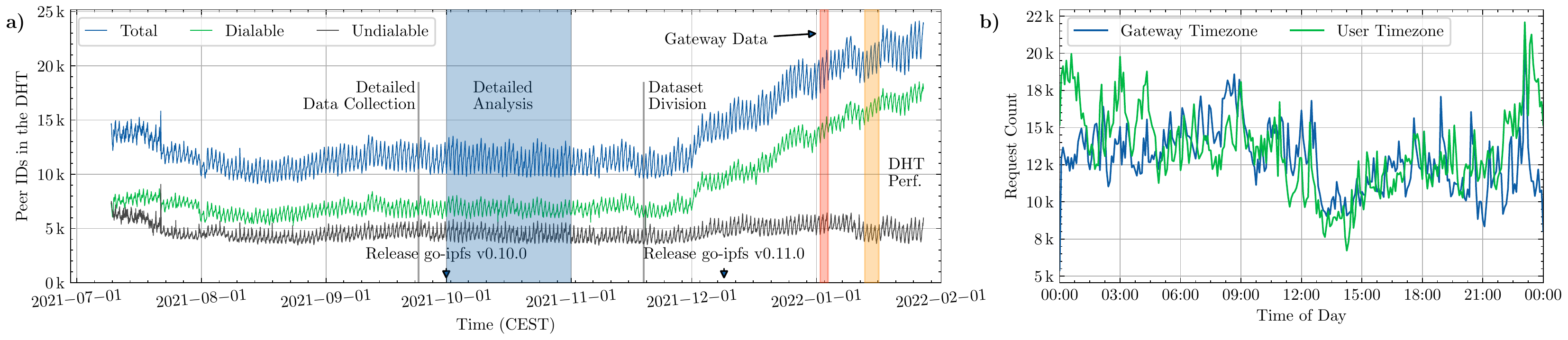}
    \caption{a) Total number of crawled peers over time and their fraction of dialable and undialable peers (one-day periodicity). The graph also depicts several events and measurement time periods which help to assess our measurement results in later sections. b) Request count of a single gateway on 2022-01-02.}
    \label{fig:plot_crawl_network_size_gateway_load}
\end{figure*}

\section{Evaluation Data}
\label{sec:evaluation_data}
We next present the data we use to evaluate IPFS as a system.
Due to its decentralized nature, it is challenging to record activities across all IPFS nodes.
Particularly, since independent node operators dominate IPFS,
no complete record exists.
To address this challenge, we compile three datasets comprising a mix of active and passive measurements. Figure \ref{fig:plot_crawl_network_size_gateway_load}a presents a time series of our measurement periods, which we describe in more detail below.

\subsection{Peer Data}
\label{sec:evaluation_data_network_crawls}
Our first dataset covers information about peers acting as DHT Servers.
As there is no central repository of such information, we employ active measurements to gather this data. 
We implement a crawler \cite{Trautwein_Nebula21} to gather a comprehensive list of all peers that are engaged in the DHT.
We run the crawler from a server in Germany every 30 minutes. 
The crawler recursively asks peers in the network for all entries in their $k$-buckets starting from the six well-known default IPFS bootstrap peers until it finds no new entries. The procedure that yields the list of all $k$-bucket peers resembles previous work~\cite{HenningsenFRS20}.

We started our measurements on 2021-07-09 and upgraded it on 2021-09-24 to collect the association of single peers with their Multiaddresses, agent version, supported protocols, and connection, handshake and crawl duration.
We then map all IP addresses to their country and Autonomous System (AS) using GeoLite2, and tag it with its CAIDA AS Rank~\cite{caida_asrank}.
In total, we have performed over \num{9500} network crawls.
Figure~\ref{fig:plot_crawl_network_size_gateway_load}a plots the number of peers we observed each day.

To quantify peer uptime, we periodically revisit all previously discovered and online peers and measure their session lengths (defined as their distinct, continuous time periods online).
Due to the scale, we adapt the probe frequency based on how often we observe a peer to be accessible. Specifically, we select an interval of 0.5x the observed uptime, starting at a minimum of 30 seconds and ending at a maximum of 15 minutes.
This is because it is more likely for peers to stay online if we observe them to have been online for an extended time period. 
We make our crawl data available for further research on IPFS with the CID:

\begin{center}
    \footnotesize
    \href{https://bafybeigkawbwjxa325rhul5vodzxb5uof73neszqe6477nilzziw5k5oj4.ipfs.dweb.link}{\texttt{bafybeigkawbwjxa325rhul5vodzxb5uof73neszqe6477nilzziw5k5oj4}}
\end{center}

\subsection{IPFS Gateway Usage Data}
\label{sec:evaluation_data_gateways}

Our second dataset covers all the \texttt{GET} requests taken from a public IPFS Gateway run by Protocol Labs (\texttt{ipfs.io}), shedding light on large-scale usage patterns of IPFS. 
This data represents a geographic subset of total gateway traffic, as the sampled traffic comes from one of several gateway instances which load balance inbound traffic using anycast.
This particular gateaway is located in the US.

The dataset covers one day of access in January 2022 with 7.1\,M user requests.
Each entry maps to a single user request and response information to the gateway comprising of request timestamp, user agent, HTTP referrer, (Maxmind-located) city, response sizes, and cache hit/miss information. 
We aggregate users by unique combinations of IP and user agent.
Overall, we find 101\,k users accessing 274\,k unique CIDs during the day of data collection. For context, Figure~\ref{fig:plot_crawl_network_size_gateway_load}b shows the number of requests received (5 minutes bins) at the gateway, based on the timezone of the gateway (Pacific Standard Time) as well as the geolocated user timezone. 
We make the access logs available on IPFS with the CID:

\begin{center}
    \footnotesize
    \href{https://bafybeiftyvcar3vh7zua3xakxkb2h5ppo4giu5f3rkpsqgcfh7n7axxnsa.ipfs.dweb.link}{\texttt{bafybeiftyvcar3vh7zua3xakxkb2h5ppo4giu5f3rkpsqgcfh7n7axxnsa}}
\end{center}

\subsection{Performance Data}
\label{sec:evaluation_data_dht_lookups}
\begin{table}[tb]
    \centering
    \footnotesize
    \caption{Number of publication and retrieval operations from each AWS region.}
    \label{tab:dht_pvd_ret_counts}
    \vspace{-0.3cm}
    \begin{tabular}{lrr}
        \hline\hline
        AWS Region                & Publications & Retrievals \\ \hline
        \texttt{af\_south\_1}     & $547$ & $2,047$ \\
        \texttt{ap\_southeast\_2} & $547$ & $2,630$ \\
        \texttt{eu\_central\_1}   & $547$ & $2,708$ \\
        \texttt{me\_south\_1}     & $547$ & $2,112$ \\
        \texttt{sa\_east\_1}      & $546$ & $2,363$ \\
        \texttt{us\_west\_1}      & $547$ & $2,704$ \\ \hline
        \textbf{Total}            & $3,281$ & $14,564$ \\
        \hline\hline
    \end{tabular}
    \vspace{-0.3cm}
\end{table}

\noindent Our third dataset focuses on benchmarking content publication and retrieval performance. 
We use six virtual machines in six different regions on AWS. Namely, the \texttt{t2.small} machines run in \texttt{me\_south\_1} (Bahrain), \texttt{ap\_southeast\_2} (Sydney), \texttt{af\_south\_1} (Cape Town), \texttt{us\_west\_1} (N. California), \texttt{eu\_central\_1} (Frankfurt) and \texttt{sa\_east\_1} (S\~ao Paulo). 
On each machine we run a \texttt{go-ipfs v0.10.0} instance acting as a DHT server node.
We use these controlled instances to interact with the public IPFS network and perform tailored performance experiments. The measurements were conducted during the shaded time period in Figure~\ref{fig:plot_crawl_network_size_gateway_load}a labeled ``DHT Perf''.

Upon each iteration, a single node announces a new $0.5\,\mathrm{MB}$ object (\ie CID) to the network.
Following this, all other nodes retrieve the object.
This involves looking up the provider and peer records, connecting to the providing peer and then downloading the object.
As soon as all remaining nodes have completed this process, they disconnect to prevent the next retrieval operation being resolved through Bitswap and instead resort to the DHT for lookup and discovery. It is worth noting that this is the closest one can get to a controlled experiment in the public IPFS network. This is because it is very difficult to replicate peers' behaviour (\eg churn, CPU and traffic load) in a simulation environment.
Table~\ref{tab:dht_pvd_ret_counts} lists the number of publications and retrievals we have performed from each region. The varying numbers of publications and retrievals stem from terminating the experiment before the instance in \texttt{sa\_east\_1} finished its latest publication and missed instructions on our control plane respectively. This does not affect the correctness of our measurement but only influences the statistical significance of the results below. The data alongside analysis code is published on IPFS with the CID:

\begin{center}
    \footnotesize
    \href{https://bafybeid7ilj4k4rq27lg45nceq4akdpetav6bcujgiym6vch5ml24tk2t4.ipfs.dweb.link}{\texttt{bafybeid7ilj4k4rq27lg45nceq4akdpetav6bcujgiym6vch5ml24tk2t4}}
\end{center}

\subsection{Ethical Considerations}
\label{sec:ethical_considerations}

The \emph{Peer} and \emph{Performance} datasets raise limited ethical concerns. They involve collecting IP addresses, yet we do not attempt to map these back to personal identities, as such analysis was not within the scope of this study. Furthermore, after performing the geolocation analysis, we have anonymized the datasets that we have made available.
The \emph{IPFS Gateway Usage Data} contains personal information, as it covers requests from web clients. This information is collected as part of our routine operations, and in line with IPFS' policies.\footnote{\url{https://docs.ipfs.io/concepts/privacy-and-encryption/}} Further, we do not trigger extra data collection. We anonymize IP addresses, and do not perform lookups on the CIDs to infer the nature of the content exchanged.

\section{Deployment Scale}
\label{sec:deployment}

We now measure the scale of the IPFS network by looking at peers, physical machines, and gateway users. Unless noted otherwise, we limit our analysis to the shaded time period labeled ``Detailed Analysis'' in Figure~\ref{fig:plot_crawl_network_size_gateway_load}a.

\subsection{Geographical Distribution}
\label{sec:geographical_distribution}

\pb{Geographical Distribution of Peers.}
Using the Peer dataset, we discover a total of \num{198964} IPFS peers (as identified by their PeerID) with \num{1998825} Multiaddresses in the DHT, covering \num{464303} unique IP addresses from 152 countries. Of these IP addresses, we were able to establish a connection to \num{253198} (\SI{54.5}{\percent}) at least once while \num{211105} (\SI{45.5}{\percent}) were always unreachable.

\begin{figure}[t]
    \centering
    \includegraphics[width=\linewidth]{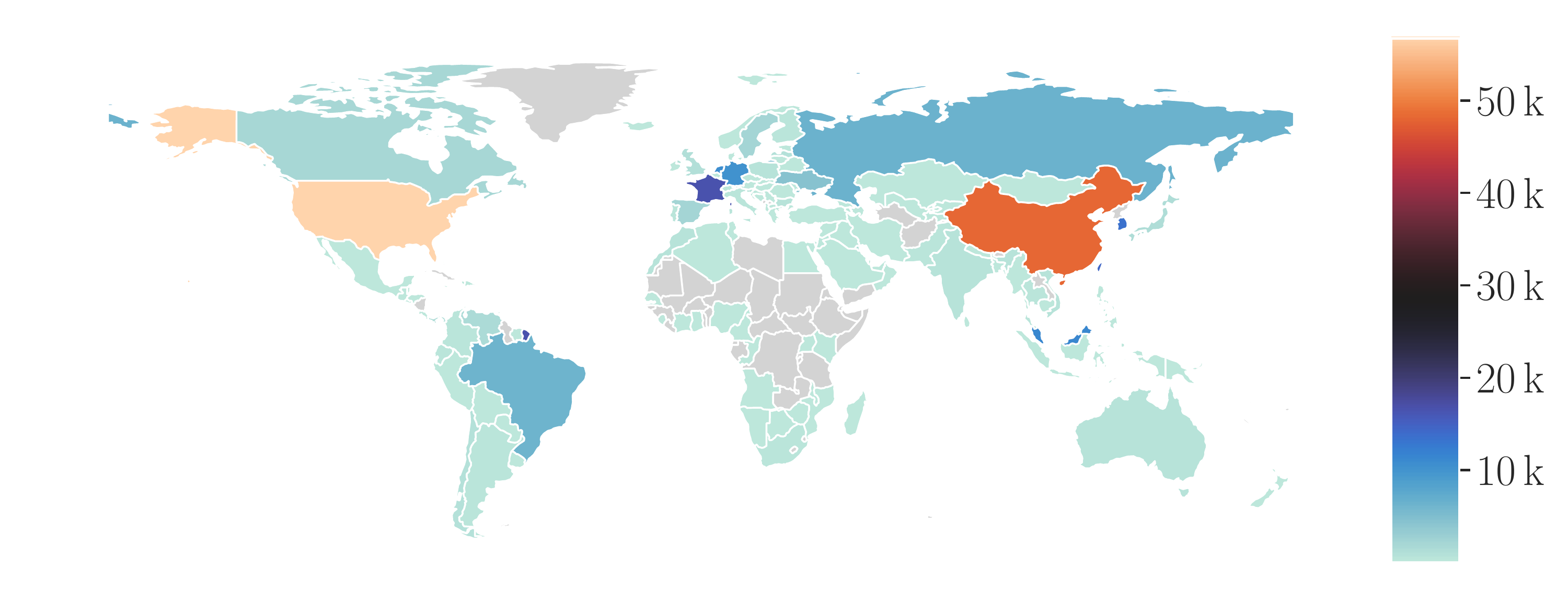} 
    \caption{Geographical distribution of peers. ``Multihoming'' peers were counted repeatedly.}
    \label{fig:crawl_geo_map_peers}
\end{figure}

Figure~\ref{fig:crawl_geo_map_peers} shows the geographical distribution of PeerIDs. Although we see widespread uptake of IPFS, we observe a high concentration in certain regions. The US (\SI{28.5}{\percent}) and China (\SI{24.2}{\percent}) dominate the share of peers, followed by France (\SI{8.3}{\percent}), Taiwan (\SI{7.2}{\percent}), and South Korea (\SI{6.7}{\percent}).
In addition, ``Multihoming'' is commonplace: around \SI{8.8}{\percent} of all peers advertise Multiaddresses that include multiple IP addresses mapped to multiple countries.

To better understand the deployment  in these different regions, Figure~\ref{fig:crawl_geo_combined}a presents the distribution of peers with \SI{>90}{\percent} uptime (``reliable''), whilst Figure \ref{fig:crawl_geo_combined}b presents the distribution of peers that were unreachable during our entire measurement period. We found \SI{1.4}{\percent} (2747) of observed peers to be ``reliable'', whereas around 1/3 of peers are never accessible. The relatively even spread of countries gives us confidence that no individual region could disrupt IPFS single-handedly, without significant strategic effort.
For example, if a single country hosted the majority of IPFS nodes, it could become possible for the state to unduly influence the wider network, which does not seem to be the case.
The distribution of reliable peers is particularly egalitarian, with the largest player (US) hosting just \SI{0.3}{\percent} of peers. We see similar trends for unreachable peers.
Although China does contain a large share (\SI{12.5}{\percent}), these unreachable peers naturally have less impact on the overall system.

Note that a single IP address can host multiple PeerIDs. Figure~\ref{fig:crawl_geo_combined}c plots a CDF of the number of peers per host.
Although the majority (\SI{92.3}{\percent}) of IP addresses host a single PeerID, we find that the top 10 IP addresses host almost \SI{66}{k} distinct PeerIDs.
This raises concerns about  potentially misbehaving peers that rotate their PeerIDs. Such peers would be able to hamper routing performance, \eg by persistently dropping requests.

\begin{figure}
    \centering
    \includegraphics[width=\linewidth]{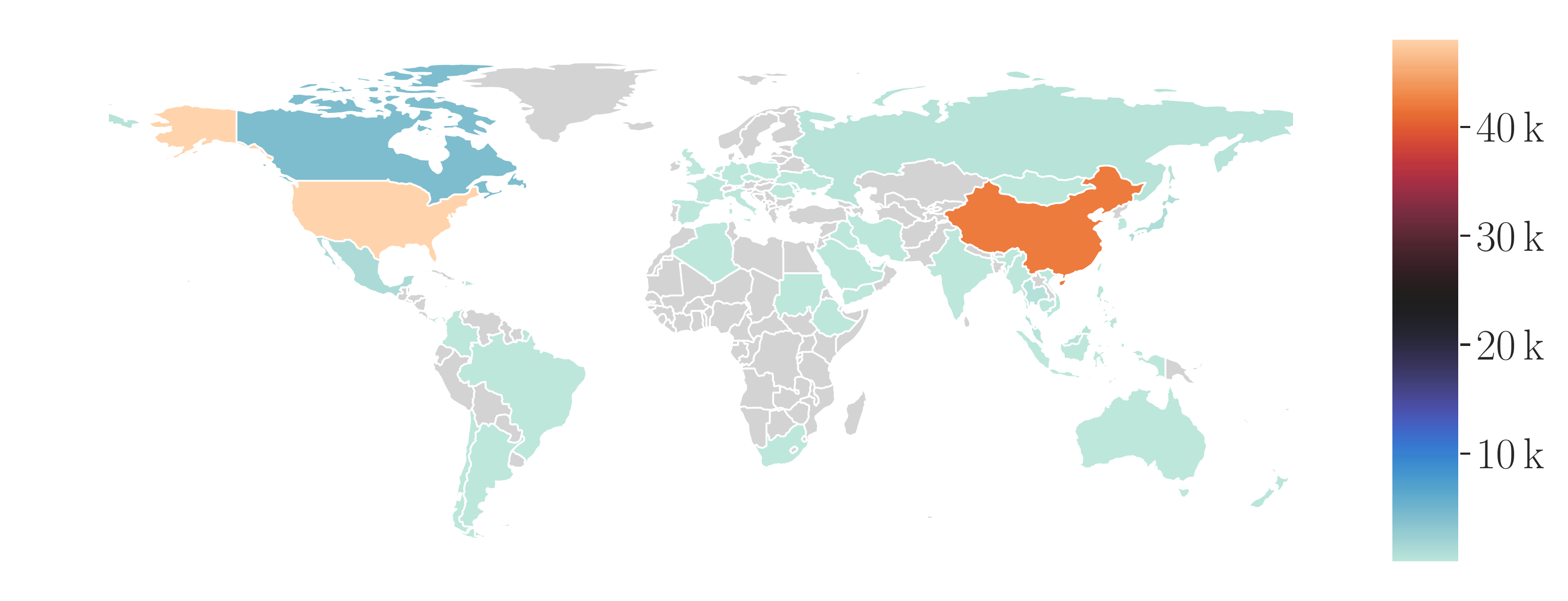} 
    \caption{Geographical distribution of users requesting content via the Gateway.}
    \label{fig:gateway_geo_map_requests}
    \vspace{-0.5cm}
\end{figure}

\pb{Geographical Distribution of Gateway Users.}
Using the gateway usage dataset, we also inspect the geographical distribution of incoming requests to get an idea of the scale of gateway usage.
In total, we observe requests from 59 countries. 
Figure \ref{fig:gateway_geo_map_requests} shows their geographical distribution.
Since the sampled node was located in the US, we find more than three-quarters of user requests to the gateway come from the US (\SI{50.4}{\percent}). This is followed by China (\SI{31.9}{\percent}), Hong Kong (\SI{6.6}{\percent}), Canada (\SI{4.6}{\percent}), and Japan (\SI{1.7}{\percent}).

\subsection{Autonomous System Distribution}
\label{sec:autonomous_system_coverage}

Next, we investigate the Autonomous System (AS) coverage of the IPFS deployment and assess potential centralization in ASes or associated cloud providers.

\begin{figure*}[ht]
    \centering
    \includegraphics[width=\textwidth]{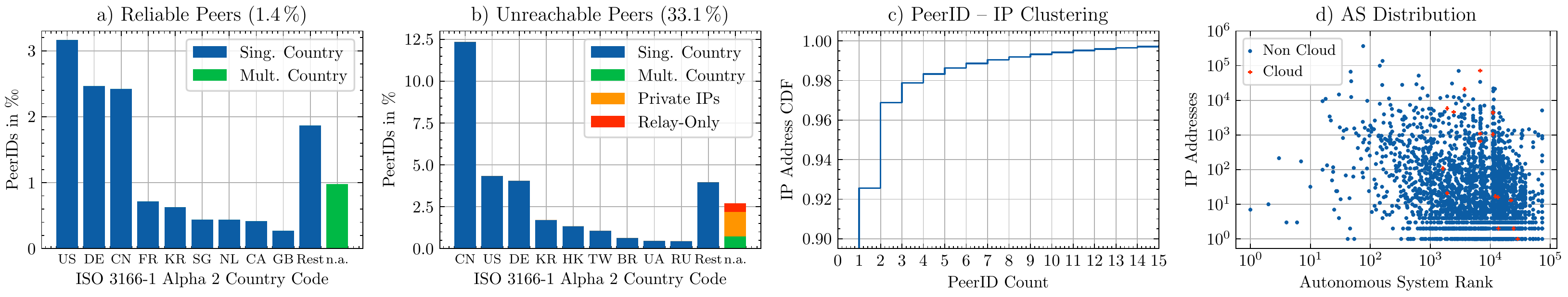} 
    \caption{a) Geographical distribution of peers that were reachable for \SI{>90}{\percent} of the measurement period. Note the \textperthousand~unit on the y-axis. b) Geographical distribution of peers that were always offline and not reachable at all during the measurement period. c) CDF of the number PeerIDs per IP address d) Distribution of IPs across ASes according to their size (measured by their AS rank).}
    \label{fig:crawl_geo_combined}
\end{figure*}

\pb{Autonomous System Coverage.}
In total, we observe peers in 2715 unique ASes.
Figure~\ref{fig:crawl_geo_combined}d presents the number of IP addresses in each AS. We rank ASes (on the x-axis) by their CAIDA AS Rank. Unsurprisingly, we see a small set of highly ranked ASes containing many IPFS hosts. 
The top 10 ASes contain \SI{64.9}{\percent} of IP addresses, whereas the top 100 contain \SI{90.6}{\percent}.
This is rather different from our prior observations related to geography, which show more egalitarian trends. Although there is a wide geographic spread, we find that each region is dominated by a small number of ASes. 
To explore this, Table~\ref{tab:asn_to_name_mapping} presents the number of IP addresses observed in the top ASes. 
Although we observe that IPFS is deployed in a large number of ASes, \SI{>50}{\percent} of IPs found are in just 5. ASes in China stand out in this regard, with two Chinese ASes containing \SI{>30}{\percent} of all observed IP addresses. This, again, raises concerns over the concentration of peers in specific networks. However, we argue that the widespread deployment across 2715 ASes, together with the decentralized structure of the P2P system (\ie with regard to indexing and resolution), would ensure resilience even in cases of severe network fragmentation, or targeted disruption attempts.

\pb{Cloud Coverage.}
Another potential cause of the above trends is the presence of prominent cloud infrastructure, used by IPFS node operators.
To this end, we use the Udger data set~\cite{udgerdb} to get a curated list of 1525 cloud providers and their IP ranges.
Table~\ref{tab:cloud_provider_coverage} lists the number of nodes that are deployed in each one. 
Contrary to expectations, we see that just a minority (\SI{<2.3}{\percent}) of IPFS nodes are hosted in cloud infrastructure.
This figure is in stark contrast to other decentralized web platforms such as Mastodon, where \SI{6}{\percent} of the infrastructure is hosted on Amazon alone~\cite{raman2019challenges}. 
This is an important finding for a decentralized storage and delivery network, and suggests that the majority of users host their own deployments.

\begin{table}
    \footnotesize
    \centering
    \caption{Autonomous systems covering \SI{>50}{\percent} of all found IP addresses.}
    \label{tab:asn_to_name_mapping}
    \vspace{-0.3cm}
    \begin{tabular*}{\linewidth}{rrrl}
        \hline\hline
        Share                & ASN & Rank    & AS Name                                              \\ \hline
        \SI{18.9}{\percent}  & 4134 & 76     & \tiny{CHINANET-BACKBONE No.31,Jin-rong Street, CN}          \\
        \SI{12.8}{\percent}  & 4837 & 160    & \tiny{CHINA169-BACKBONE CHINA UNICOM China169 Back., CN}    \\
        \SI{9.6}{\percent}   & 4760 & 2976   & \tiny{HKTIMS-AP HKT Limited, HK}                            \\
        \SI{6.9}{\percent}   & 26599 & 6797  & \tiny{TELEFONICA BRASIL S.A, BR}                            \\
        \SI{5.3}{\percent}   & 3462 & 340    & \tiny{HINET Data Communication Business Group, TW}          \\
        \hline\hline
    \end{tabular*}
    \vspace{0.3cm}
    \caption{Percentage of nodes hosted on cloud providers. The table shows the top ten and selected cloud providers.}
    \label{tab:cloud_provider_coverage}
    \vspace{-0.3cm}
    \begin{tabular*}{\linewidth}{ll@{\extracolsep{\fill}}rr}
        \hline\hline
        Rank & Provider                     & IP Addresses & IP Address Share    \\ \hline
        1 & Contabo GmbH                 & \num{2038}  & \SI{0.44}{\percent}  \\
        2 & Amazon AWS                   & \num{1792}   & \SI{0.39}{\percent} \\
        3 & Microsoft Azure/Coporation   & \num{1536}   & \SI{0.33}{\percent} \\
        4 & Digital Ocean                & \num{836}    & \SI{0.18}{\percent} \\
        5 & Hetzner Online               & \num{592}    & \SI{0.13}{\percent} \\
        6 & GZ Systems                   & \num{346}    & \SI{<0.10}{\percent} \\
        7 & OVH                          & \num{341}    & \SI{<0.10}{\percent} \\
        8 & Google Cloud                 & \num{286}    & \SI{<0.10}{\percent} \\
        9 & Tencent Cloud                & \num{258}    & \SI{<0.10}{\percent} \\
        10 & Choopa, LLC. Cloud           & \num{244}    & \SI{<0.10}{\percent} \\
        12 & Alibaba Cloud                & \num{180}    & \SI{<0.10}{\percent} \\
        13 & CloudFlare Inc               & \num{140}    & \SI{<0.10}{\percent} \\
        27 & Oracle Cloud                 & \num{27}     & \SI{<0.10}{\percent} \\
        54 & IBM Cloud                    & \num{9}      & \SI{<0.10}{\percent} \\ \hline
          & 235 Other Cloud Providers      & \num{2017}   & \SI{0.43}{\percent} \\ 
           & Non-Cloud                    & \num{453661} & \SI{97.71}{\percent}\\
        \hline\hline
    \end{tabular*}
    \vspace{-0.3cm}
\end{table}

\subsection{Churn}
\label{sec:churn}

\begin{figure}[h]
    \centering
    \includegraphics[width=\linewidth]{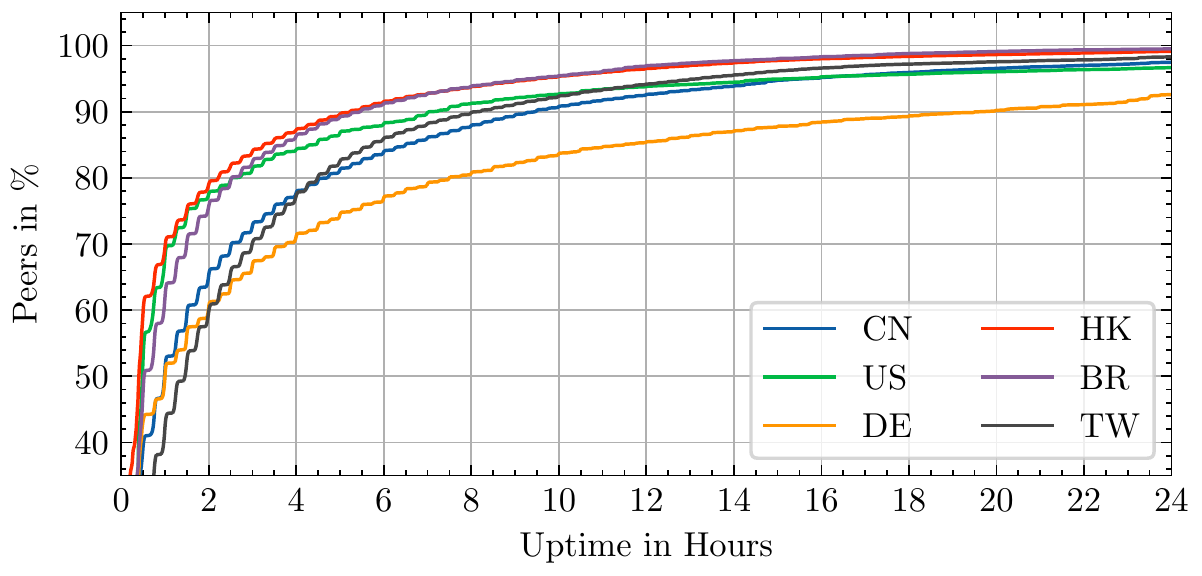} 
     \caption{Churn rate by region for \num{467134} session observations. Each line shows the cumulative distribution function of the DHT peer uptimes in a given region.
     The step shape correlates with the sampling interval of our crawler.}
    \label{fig:crawl_churn_combined}
\end{figure}




Churn refers to the action of peers arriving and departing from the network. The churn rate within a system like IPFS is important as it influences decisions such as for how long should peers retain information about other (online) peers.
To calculate churn from our Peer dataset, we follow the method used in \cite{StutzbachR06, SaroiuGG, RoselliLA} for long session handling to minimize bias towards shorter sessions and account for peers that stay online beyond our selected measurement time window.
Figure \ref{fig:crawl_churn_combined} plots CDFs of DHT peer uptimes for countries that stood out in the deployment analysis of Section~\ref{sec:deployment}, based on \num{467134} session observations that started in the first half of our ``Detailed Analysis'' time window (see Figure~\ref{fig:plot_crawl_network_size_gateway_load}a).

We observe that uptime tends to be short, with \SI{87.6}{\percent} of sessions under 8 hours and only \SI{2.5}{\percent} of sessions exceeding 24 hours. This helps explain our design decision to replicate records on a relatively large number of peers ($k=20$, see Section~\ref{sec:dht}). Briefly, we also note that stability varies greatly based on region. For example, whereas the median uptime for Hong Kong is just \SI{24.2}{\minute}, it is more than double that figure for Germany.

\subsection{Discussion \& Takeaways}
\label{sec:deployment_takeaways}

IPFS is designed to be highly decentralized. Although we observe geographical agglomeration in certain regions, we also see IPFS being widely adopted around the globe. Especially important is the finding that fewer than \SI{2.3}{\percent} of IPFS nodes run in major cloud platforms.
This suggests that individuals are running IPFS nodes on personal or on-premises commodity hardware. As a downside, this contributes to the high churn rate observed (only \SI{2.5}{\percent} of peers stay online for more than \SI{24}{\hour}). Further, although we find peers in 2715 ASes, the top 10 contain 64.9\% of peers alone. This suggests that we should push harder to broaden deployment and avoid consolidation in a minority of ASes.


\begin{figure*}[h]
    \centering
    \includegraphics[width=\textwidth]{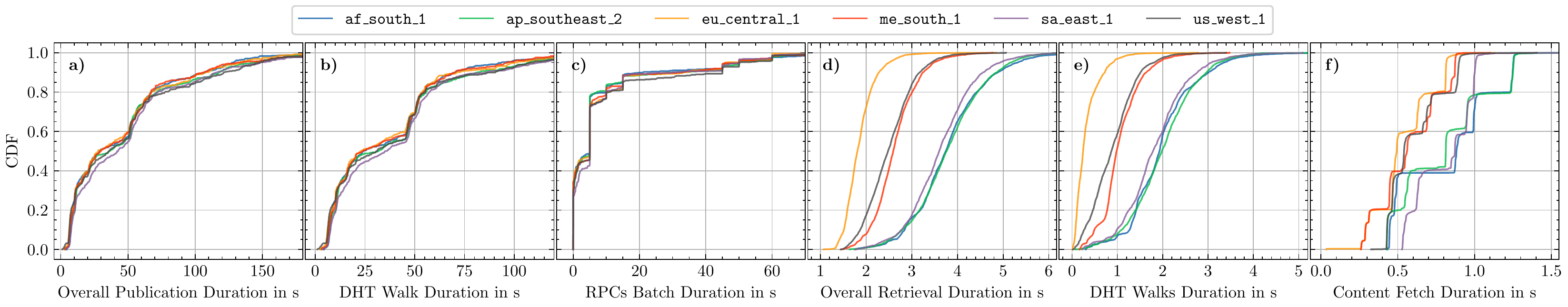}
    \caption{CDFs for content publication (a--c) and content retrieval (d--f) for each AWS region that we employed for our lookup measurements. a) The overall publication duration  includes b) walking the DHT, and c) the batch of RPCs to store the provider record at suitable peers. 
    d) The overall content retrieval duration  includes e) walking the DHT twice, and 
    f) fetching the data from the providing peer. The sample size is \num{4324} in all graphs across all AWS regions combined.}
    \label{fig:dht_pvdret_process_complete}
\end{figure*}

\section{IPFS Performance Evaluation}
\label{sec:evaluation}

In this section, we evaluate the performance of IPFS' two core functions: content publication and content retrieval. We also turn to our Gateway dataset to better understand IPFS's usage through browsers.

\subsection{Content Publication Performance}
\label{sec:evaluation_content_publication_performance}


As explained in Section \ref{sec:content-publication}, the content publication process consists of two steps. First, the content is imported into the local IPFS node and second, provider records are stored with suitable peers. The first (\ie importing content) step was covered in \cite{LajamH21}. Here, we focus on the publication of provider records to the network.

\pb{Overall Delay.}
Figure~\ref{fig:dht_pvdret_process_complete}a shows CDFs for the duration of the overall content publication process. We present separate results for each AWS region from which the measurements are launched. The overall publication process across all regions takes \SI{33.8}{\second}, \SI{112.3}{\second}, and \SI{138.1}{\second} in the 50th, 90th, and 95th percentiles, respectively. Note that the publication delay is independent of the content size as only the provider record is being published. Table~\ref{tab:dht_performance_percentiles} breaks down the publication duration percentiles for each AWS region individually. We see similar results across all regions despite the larger build-up of presence in the US and China.

\pb{Delay Breakdown.}
The two main constituents of delay are the DHT walk (to discover the 20 closest peers to the CID) and the RPC operations that push provider records to the chosen peers. 
Figure~\ref{fig:dht_pvdret_process_complete}b presents CDFs of the DHT walk duration. We see that this constitutes the largest component of the delay. On average, the DHT walk covers \SI{87.9}{\percent} of the overall delay. Naturally, optimizing this is a target of our future work.

In contrast, Figure \ref{fig:dht_pvdret_process_complete}c shows the CDFs for the duration of the batch of RPCs. This reveals much faster performance, with \SI{43.3}{\percent} of RPC batches completing in under \SI{2}{\second}.
That said, \SI{53.7}{\percent} exceed \SI{5}{\second}, and \SI{11.3}{\percent} exceed \SI{20}{\second}.
Closer inspection reveals a number of spikes in the distribution. These are driven by various timeout operations in the protocol. 
For example, the spike at \SI{5}{\second} is caused by dial timeouts on the transport level of the TCP and QUIC implementations, whereas the spike at \SI{45}{\second} is caused by the handshake timeout of the Websocket transport.
These timeouts stem from less responsive peers.
This points to unavoidable challenges when operating decentralized infrastructure, where the responsiveness of peers can be difficult to predict.

\begin{table}
    \centering
    \footnotesize
    \caption{Latency percentiles of the overall DHT publication and retrieval operations from different AWS regions.}
    \label{tab:dht_performance_percentiles}
    \vspace{-0.3cm}
    \begin{tabular*}{\linewidth}{l@{\extracolsep{\fill}}|rrr|rrr}
    \hline\hline
        \multicolumn{1}{c}{}        &          \multicolumn{3}{c}{Publication Percentiles}              &         \multicolumn{3}{c}{Retrieval Percentiles}            \\
        AWS Region                  & 50th                & 90th                 & 95th                 & 50th               & 90th              & 95th            \\ \hline
        \texttt{af\_south\_1}       & \SI{28.93}{\second} & \SI{107.14}{\second} & \SI{127.22}{\second} & \SI{3.75}{\second} & \SI{4.88}{\second} & \SI{5.31}{\second} \\
        \texttt{ap\_southeast\_2}   & \SI{36.26}{\second} & \SI{117.74}{\second} & \SI{142.79}{\second} & \SI{3.76}{\second} & \SI{4.85}{\second} & \SI{5.15}{\second} \\
        \texttt{eu\_central\_1}     & \SI{27.70}{\second} & \SI{106.91}{\second} & \SI{133.27}{\second} & \SI{1.81}{\second} & \SI{2.28}{\second} & \SI{2.50}{\second} \\
        \texttt{me\_south\_1}       & \SI{29.32}{\second} & \SI{105.45}{\second} & \SI{130.48}{\second} & \SI{2.59}{\second} & \SI{3.24}{\second} & \SI{3.48}{\second} \\
        \texttt{sa\_east\_1}        & \SI{42.32}{\second} & \SI{115.45}{\second} & \SI{148.04}{\second} & \SI{3.60}{\second} & \SI{4.56}{\second} & \SI{4.93}{\second} \\
        \texttt{us\_west\_1}        & \SI{36.02}{\second} & \SI{121.13}{\second} & \SI{147.59}{\second} & \SI{2.48}{\second} & \SI{3.17}{\second} & \SI{3.42}{\second} \\
    \hline\hline
    \end{tabular*}
\end{table}

\subsection{Content Retrieval Performance}
\label{sec:evaluation_content_retrieval_performance}


\pb{Overall delay.}
Figure~\ref{fig:dht_pvdret_process_complete}d shows CDFs for the duration of the overall retrieval operation. 
We observe success rate of \SI{100}{\percent}, confirming the reliability of the system. 
Although we control the providing and retrieving peers for this test, this high success rate is not self-evident since retrieval operations involve many interactions with peers outside our control.
However, we experience notable performance diversity, with delays on average higher than typical web page loading times.
Overall, retrieval performance is much faster than publication. 
The content retrieval process across all regions takes \SI{2.90}{\second}, \SI{4.34}{\second}, and \SI{4.74}{\second} in the 50th, 90th, and 95th percentiles, respectively.
We emphasize that, in contrast to HTTP, this includes the \emph{lookup} time, whereby CIDs are mapped to eligible locations to obtain the content from (in contrast to HTTP URLs that encode the location a priori). 

Table~\ref{tab:dht_performance_percentiles} further breaks down the retrieval duration percentiles for each AWS region individually.
Interestingly, we observe variations across regions. For example, the median retrieval delay in central Europe is just \SI{1.81}{\second}, compared to \SI{3.75}{\second} from South Africa. These slower retrievals are in-line with prior measurements of web performance in Africa~\cite{fanou2016pushing}.
Further, recall that due to Bitswap, DHT queries are only launched after a timeout of 1s (see Section~\ref{sec:content-routing}), setting a minimum baseline of 1s delay. Finding intelligent ways to minimize this initial delay is a key line of future work.

\pb{Delay Breakdown.}
The three main constituents of retrieval delay are \one~the opportunistic content discovery through Bitswap, which (if unsuccessful) times out at \SI{1}{\second}\footnote{At this point, it is also worth highlighting that due to our experiment setup, none of the requests are satisfied at the Bitswap level, hence, in all experiments presented here, retrievals include an extra \SI{1}{\second} for the Bitswap timeout and constitute worst-case scenarios.}; \two~the time spent walking the DHT to find provider and peer records; and \three~the content exchange operation to fetch the object. Figure~\ref{fig:dht_pvdret_process_complete}e presents CDFs for the combination of DHT walks to discover provider and peer records. The median duration across all regions for a single DHT walk is \SI{622}{\milli\second} and, therefore, significantly faster than the DHT walk for the publication operation. This is because a retrieval DHT walk terminates after the discovery of a single record-hosting node, rather than 20 in the case of publication. More generally, it is positive to see that single DHT walks complete with sub-second latency for most regions. In fact, we see that for all regions both DHT walks (first to find the provider record and second to find the peer record) complete in less than \SI{2}{\second} for \SI{50}{\percent} of the retrievals. This is in stark contrast to previous DHT deployments, where latency can even exceed a minute~\cite{crosby2007analysis}. 

Finally, Figure~\ref{fig:dht_pvdret_process_complete}f shows the content fetch duration. This includes peer routing and content download (see Section~\ref{sec:content-routing}).
The figure follows a step-wise pattern, with each step corresponding to a unique combination of nodes publishing and fetching the content (recall that each node is in a different region). Over \SI{99}{\percent} of all content exchange operations take less than \SI{1.26}{\second}. We emphasize that this number is dependent on the amount of data being exchanged (which is set to \SI{0.5}{MB} in our measurement setup), \ie higher data volume reduces the overall retrieval percentage spent on discovery.

\pb{Retrieval Stretch.}
\begin{figure}
    \centering
    \includegraphics[width=\linewidth]{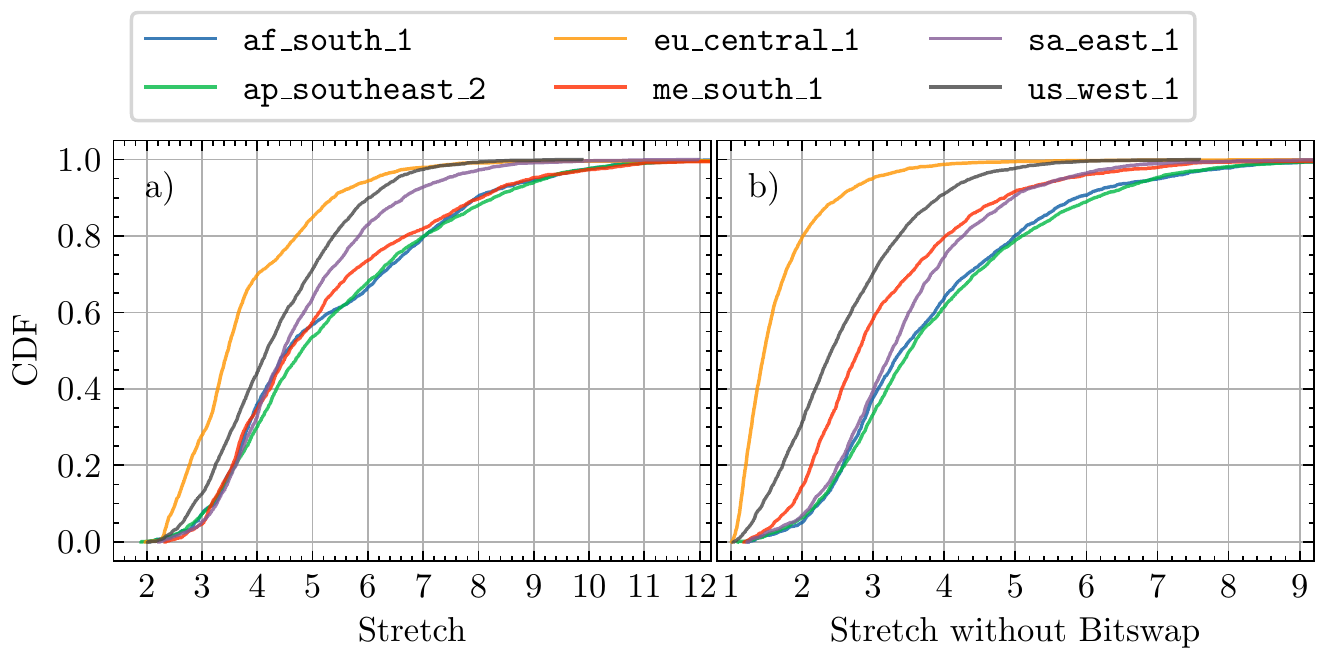}
    \caption{CDFs of the retrieval stretch (defined in equation~\ref{eq:definition_stretch})
    for each AWS vantage point with (a) and without (b) the initial Bitswap timeout.}
    \label{fig:dht_ret_stretch}
\end{figure}
To compare IPFS retrieval operations against HTTPS, we measure \textit{request stretch} as the ratio of the IPFS retrieval time \vs the \textit{estimated} HTTPS retrieval time (\eg retrieving an object using a URL via HTTPS):
\begin{align}
    \mathrm{Stretch} &= \frac{\text{IPFS Content Retrieval Time}}{\text{HTTPS Content Retrieval Time}} \\[0.2cm]
    & = \frac{\text{Discover} + \text{Dial} + \text{Negotiate} + \text{Fetch}}{\text{Dial} + \text{Negotiate} + \text{Fetch}}.
    \label{eq:definition_stretch}
\end{align}
In IPFS vernacular, Dial is equivalent to the TCP handshake; Negotiate is equivalent to the TLS handshake; and Fetch is equivalent to issuing the \texttt{GET} request and byte transfer.
Put simply, we roughly estimate the ``HTTPS Content Retrieval Time" by subtracting the DHT ``Discover" latency from the ``IPFS Content Retrieval Time". 
Figure~\ref{fig:dht_ret_stretch}a shows CDFs for the stretch of all content retrieval operations for each AWS region. The majority of content retrieval operations take at least four times as long as the equivalent HTTPS request across all AWS regions.
This could be characterized as the ``cost of decentralization'' when using IPFS.
However, recall that the ``Discover'' step in the IPFS case also includes the Bitswap delay mentioned in Section~\ref{sec:content-routing}, which adds 1s and is always present in our experiments, due to our experimental setup.
To inspect the impact of this, Figure~\ref{fig:dht_ret_stretch}b shows the stretch CDFs without the initial Bitswap delay. We note that especially for the IPFS instance in central Europe, the ``cost of decentralization'' is comparably low with a stretch $<2$ for \SI{80}{\percent} of content retrieval operations. This shows that, arguably, running DHT lookups in parallel to Bitswap could be superior, by trading additional network requests for faster retrieval times.

\begin{figure}
    \centering
    \includegraphics[width=\linewidth]{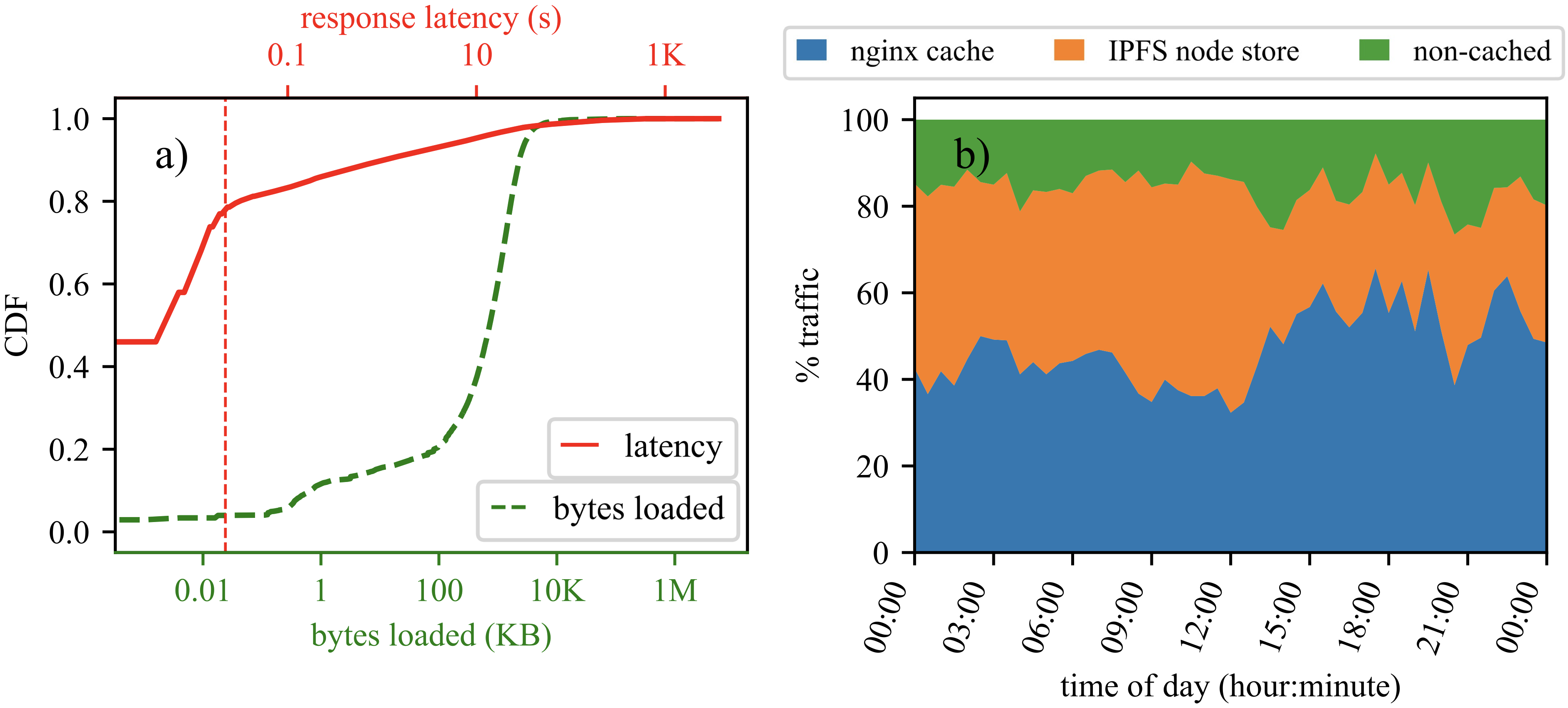} 
    \caption{(a) Distribution of upstream response latency at gateway (top $x$-axis) and bytes downloaded per request to the gateway from peers (bottom $x$-axis). Latencies incurred for serving non-cached requests are present to the right of dotted red line. (b) Proportion of cached and non-cached traffic binned at every \SI{30}{\minute}.}
    \label{fig:gateway_response_time}
\end{figure}
\begin{table}[]
\caption{Traffic and latencies to the gateway, when content is served from default nginx cache, the IPFS node store, and from other peers in the IPFS network.}
\label{tab:cache_performance}
\resizebox{\columnwidth}{!}{%
\begin{tabular}{lrrr}
\hline\hline
                & nginx cache & IPFS node store & Non Cached \\
\midrule
Latency (Median) & \SI{0}{\second}      & \SI{8}{\milli\second}    & \SI{4.04}{\second}      \\
Traffic Served  & \SI{46.4}{\percent} & \SI{38.0}{\percent} & \SI{15.6}{\percent}    \\
Requests Served & \SI{46.0}{\percent} & \SI{40.2}{\percent}   & \SI{13.8}{\percent}    \\
\hline\hline
\end{tabular}
}
\end{table}
\subsection{Gateway Evaluation}
\label{sec:overhead}

We finally turn to the Gateway data to inspect its performance, as well as usage.

\pb{Object Size.} 
The gateway dataset allows us to obtain a sample of content objects being requested from clients. This provides vantage on the size distribution. In total, we observe 6.57 TB being downloaded during our measurement period. Figure~\ref{fig:gateway_response_time}a ($x_1$-axis) presents the distribution of object sizes (bytes loaded). 
We see a wide variety. The majority (79.1\,\%) are above 100\,KB, with a median of 664.59\,KB. 
We also find there is no correlation between the object size and the latency (Pearson correlation coefficient of 0.13), again confirming that delays are primarily driven by size-agnostic operations (\eg DHT walks).

\pb{Retrieval Delays.}
The gateway further provides a 
second source of IPFS retrieval latency measurements. Each \texttt{GET} request log entry is accompanied by the delay it took the gateway to retrieve the object (either from IPFS or its cache).
Figure~\ref{fig:gateway_response_time}a ($x_2$-axis) plots the distribution of retrieval latencies (as defined in Section~\ref{sec:evaluation_content_retrieval_performance}).

46\,\% of fetches have a retrieval delay of 0, which indicates a default nginx cache hit.
The remaining 54\,\% enter IPFS via the peer bridge co-located on the gateway.
67.1\,\% of these are served from the local IPFS node store itself, resulting consistently in a delay below 24ms (as marked by the vertical line in the figure).
Recall, this store contains content manually uploaded by the Web3 and NFT Storage Initiatives.
Due to the efficacy of this caching, 76\,\% of requests are delivered in less than \SI{250}{\milli\second}.
This is substantially better than observed from our prior measurements (see Section~\ref{sec:evaluation_content_retrieval_performance}) and exemplifies a benefit of aggregating demand at the gateways.

\pb{Cache Hit Rates.}
Figure~\ref{fig:gateway_response_time}b presents the fraction of cache hits \vs misses across the one day period in our dataset.
Table~\ref{tab:cache_performance} summarizes the key performance results. Hit rates in the nginx cache vary from 32.3\,\% (at 00:00) to 65.6\,\% (at 17:30). 
These hits serve close to 50\,\% of the total traffic.
When combined with pinned content stored in the gateway's local IPFS node, hit rates exceed 80\,\%.
This indicates that the gateway offloads substantial traffic from the remaining IPFS network and (as shown in Table~\ref{tab:cache_performance}) reduces fetch times.
That said, content that is not cached experiences slower retrieval times, although this only constitutes 15.6\,\% of bytes transferred.
This suggests that augmenting IPFS with a gateway model \emph{does} offer a meaningful strategy for reducing delays by aggregating demand via the cache.

\pb{Gateway Referrals.}
We finally inspect the HTTP referrals observed in the gateway's nginx logs. These indicate the website that redirected the web client to the gateway. 
Indeed, the majority of this traffic (\SI{51.8}{\percent}) is referred by third party websites, confirming the integration of IPFS with more traditional HTTP hosted sites. Interestingly, \SI{70.6}{\percent} of this referred traffic belongs to just 72 semi-popular websites (rank \SIrange{10}{50}{k} based on Tranco list\footnote{List collected a day after the log capture at the gateway}~\cite{LePochat2019tranco}). 
The majority of these parent sites are hosted in the US (\SI{47.3}{\percent}), Iceland (\SI{20.0}{\percent}) and Canada (\SI{12.7}{\percent}). 
Manual inspection shows that these are primarily video streaming and Non-Fungible Token (NFT) related sites, as many people have chosen to store NFTs on IPFS.\footnote{\url{https://docs.ipfs.io/how-to/best-practices-for-nft-data/}} This suggests that IPFS is receiving widespread uptake in these targeted use cases. 

\subsection{Discussion \& Takeaways}
\label{sec:evaluation_discussion_key_takeaways}

IPFS is designed to offer publication and retrieval delays capable of supporting a range of applications. 
IPFS attains shorter retrieval delays compared to prior large-scale DHT deployments.
For example, the median lookup latency for the Kademlia implementation in BitTorrent (both Mainline and Azureus) exceeds a minute~\cite{crosby2007analysis}, due to excessive dead nodes, as well as slower links at that time. Our observations indicate that the distinction between server and client peers (see Section~\ref{sec:dht}) (after the v0.5 release of IPFS) has given a significant boost to the performance of IPFS, as peers avoid costly operations of attempting to punch through NATs, failing and timing out eventually.

In contrast, for half of our probes, the record lookup stage takes under one second, showing a dramatic improvement over past DHT deployments. 
That said, the overall IPFS content retrievals have a median stretch of $4.3$ (compared to HTTPS), although we find that  removing the initial Bitswap timeout can decrease the stretch to $<2$ for 80\,\% of retrievals in well-connected areas.
We argue that this makes IPFS suitable for various applications, including video on demand, file sharing and other social networking services. 
In comparison, publication delays are higher (median of \SI{33.8}{\second}), due to the need to push records onto 20 distinct peers (to protect from churn, see Section~\ref{sec:churn}). 
For many applications, this delay would be acceptable (\eg publishing a movie), as exemplified by the number of applications already built on IPFS (see~\cite{ipfsecosystem} for a non-exhaustive list).
However, we acknowledge that the performance users experience over the last mile to their homes will differ from what we have observed from AWS data centres.

We have further experimented with several ways to streamline uptake and performance, including the use of gateways. Although arguably a centralized entity, we find that it can substantially speed-up retrievals with 46\,\% of retrievals being served via the nginx cache. Uptake is wide, with 101\,k users hitting the gateway in a single day. 
We further note that anybody can set up their own gateway, ensuring that no single entity could emerge as a gatekeeper.

\section{Related Work}
\label{sec:related-work}


\pb{P2P Networks.}
There have been countless P2P overlay architecture proposals, including dozens of DHT structures~\cite{stoica2001chord,zhao2004tapestry,kaashoek2003koorde,rowstron2001pastry,10.1145/1851182.1851198}, and tens of applications, \eg large-scale content delivery platforms~\cite{cohen2003incentives}, and 
services such as decentralized social networks~\cite{graffi2011lifesocial}.
Rather than devising an entirely new system, IPFS utilizes the Kademlia DHT for content indexing~\cite{maymounkov2002kademlia}. 
There have been various attempts to optimize the performance and usability of such DHTs via caching~\cite{saleh2006modeling}, network-aware peer selection~\cite{kaune2009modelling}, and parallelizing lookups~\cite{stutzbach2006improving}.
IPFS builds on these and currently constitutes one of the largest deployments of peer-to-peer networks ``in the wild'' (alongside BitTorrent, which also uses Kademlia~\cite{cohen2003incentives}).
IPFS also strives to be censorship resistant. Approaches such as 
Freenet~\cite{clarke2001freenet} and Wuala~\cite{mager2012measurement} have similar goals. These work by storing encrypted content across an arbitrary subset of peers.
In contrast, IPFS takes a BitTorrent-like approach where nodes store only the content they are interested in. 



\pb{Evaluation of Operational DHTs.} 
Closer to our work are studies that measure operational DHTs~\cite{10.1145/1298306.1298325, 6688697}. 
For example, the authors in \cite{10.1145/1298306.1298325} report churn rates similar to our findings. However, they have not carried out similar controlled experiments from multiple vantage points. Instead, they attempted DHT lookups and report latencies in the order of tens of seconds, significantly slower than IPFS achieves.
The authors in \cite{crosby2007analysis} and \cite{267312} measured performance in the BitTorrent implementation of Kademlia.
Their results contrast starkly with our own, showing a substantial number of failed nodes that negatively impact lookup times. Finally, Stutzbach and Rejaie~\cite{stutzbach2006improving}, modeled Kademlia performance and proposed a set of improvements. Despite this, both studies revealed substantially worse performance than attained by IPFS.


\pb{The Fediverse.}
The growth of the IPFS network has occurred in tandem with other Decentralized Web technologies, most notably the ``fediverse''. The fediverse includes a number of server-based federated services, \eg Mastodon~\cite{raman2019challenges}, Pleroma~\cite{hassan2021exploring} and Diaspora~\cite{guidi2018managing}.
Closest to our own work is Nextcloud, which provides a federated file storage platform, with IPFS integration in addition to server-local storage \cite{nextcloud_files_external_ipfs}.
This is complementary to our own work, and operates in a similar fashion to IPFS gateways. In contrast, however, IPFS can continue to operate without gateways, whereas fediverse apps are entirely dependent on the uptime of the federated servers (see~\cite{raman2019challenges} for detailed analysis).


\pb{Incentives.}
There have been several studies that look at incentivizing participation in P2P systems~\cite{202577}. There are also several large operational decentralized and incentivized P2P storage networks~\cite{filecoin-whitepaper,sia-whitepaper,arweave-whitepaper} with Filecoin ~\cite{filecoin-whitepaper} in particular building directly on top of IPFS and being the largest decentralized and incentivized storage network at the time of writing~\cite{storageindex}.
IPFS does not incentivize data storage, sharing, or participation in the network. IPFS can be considered a best-effort caching, storage and distribution layer that sits underneath the incentive structures layer.
Section~\ref{sec:deployment} shows that even without direct incentives, participation in IPFS is widespread, in-part because nodes are not required to store other users' content.

In order to go a step further and provide availability guarantees, ``pinning services" offer to host and provide user content for a fee (see Pinata, or Infura). Availability guarantees can also be provided by Filecoin \cite{filecoin-whitepaper}. Filecoin's total storage capacity stands at 17.5 EiB at the time of writing (compared to AWS's 194 EiB\footnote{\url{https://www.storageindex.io/}}). Given that Filecoin is built on top of IPFS, it is expected to increase traffic in the IPFS network significantly in the coming years.


\pb{Content-Based Addressing.} 
Content-based addressing has been widely used in P2P networks for many years~\cite{10.1145/964723.383072,li2008adaptive,carzaniga2004routing}. More recently, content addressing has received significant uptake by the Information-Centric Networking (ICN) community. 
Prominent architectural proposals in this field include Networking Named Content (NNC) \cite{10.1145/1658939.1658941}, Named Data Networking (NDN) \cite{10.1145/2656877.2656887}, NetInf (which is also DHT-based) \cite{10.1016/j.comcom.2013.01.009}, Curling~\cite{5723808}, as well as Secure Scuttlebutt (SSB) \cite{10.1145/3357150.3357396}, more recently.
IPFS is complementary to these proposals. Whereas they largely introduce content-based addressing at the network-layer, IPFS relies on application-layer routing. Clearly, if designed carefully, synergies between network-layer and overlay-based approaches can complement each other and result in a universal content-based addressing networking stack.

\pb{Decentralized Web Data Management.}
Researchers have also looked at data management and decentralized content storage more generally~\cite{taheri2015security,schwittmann2013sonet}.
For example, various projects have attempted to decentralize data control, \eg DataBox~\cite{perera2017valorising}, SOLID~\cite{mansour2016demonstration}, and SocialGate~\cite{koll2017socialgate}. 
These operate local datastores for individual users, \eg running on a physical home appliance. 
While these solutions focus on controlling data usage and access,
IPFS has a broader focus in providing a decentralized storage system, \eg by serving as a back-end~\cite{9472772}.



\section{Conclusions}
\label{sec:conclusion}

This paper has detailed the design, implementation and deployment of the InterPlanetary File System (IPFS).
As well as presenting the core design of IPFS, we have presented measurement tooling that allows us to gain vantage on its decentralized operations. 
We have shown that IPFS has received widespread uptake, covering 152 countries and 2715 ASes.
This uptake is in-part enabled by our hybrid gateway design, which has  received considerable usage.
The IPFS codebase\footnote{\url{https://github.com/ipfs/ipfs}} and datasets are freely available.

There are a number of avenues of future work.
Most relevant, we plan to focus on minimizing retrieval and publication latency.
For instance, we plan to exploit the Bitswap protocol to preemptively pair peers who may have similar content interests. 
We also plan to continue our large-scale performance monitoring to gain greater longitudinal insight into the scale and performance of the several components of the IPFS architecture. We plan to expand our studies to components such as the Hydra boosters,\footnote{\url{https://github.com/libp2p/hydra-booster}} which we have not covered here due to space constraints and their limited adoption.

We further wish to emphasize that IPFS offers properties that go beyond those discussed within the remit of this paper.
IPFS is an open, permissionless system and, as such, moderation remains a challenge.
Studying potential forms of misuse is therefore a key line of future work. 
For instance, there have been reports that the Storm botnet\footnote{\url{https://en.wikipedia.org/wiki/Storm_botnet}} lives on the IPFS network, but the precise intention and activities of the botnet have not yet been identified. Our initial monitoring did not show abusive activity by IPStorm nodes (\eg frequent switching of PeerIDs, higher than usual rate of churn, or unresponsive behaviour), but investigation is part of our future plan.
Similarly, we are yet to evaluate the resilience of IPFS or its capacity to sustain various types of information attacks (\eg censorship). Although our results shed light on these matters, for example, revealing that IPFS is robust to churn and is sufficiently geographically distributed to avoid single points of failure, this area is ripe for further work.
As part of this, it is important to better understand the reliance that IPFS has on other centralized infrastructure (\eg cloud platforms), and to explore the types of networks that host IPFS nodes (\eg homes, universities).
Thus, we plan to continue developing measurement techniques that can quantify these issues, and feedback into our ongoing community-driven design process.

\section*{Acknowledgements}
We would like to acknowledge and thank the numerous contributors to IPFS, without whom this paper would not be possible.
We would like to particularly thank Adin Schmahmann, the core maintainer of the go-ipfs codebase and mastermind of several of the techniques discussed in this paper, and Petar Maymounkov, the original author of Kademlia and also a maintainer of go-ipfs.
Both generously provided endless discussions and explanations around the concepts presented here as well as offering thoughtful inputs into our methodologies. We would also like to thank the Pegasys Team (part of Consensys) and especially Zhenyang Shi for developing the basis for some of the measurement tools used here. This work was part-funded by projects 
EPSRC REPHRAIN “Moderation in Decentralised Social Networks”,
EP/S033564/1,
EP/W032473/1 and EU Horizon 2020 grant agreements No 871793 (Accordion), No 101016509 (Charity). Gareth Tyson is corresponding author.

\bibliographystyle{ACM-Reference-Format}
\bibliography{ipfs-arch.bib}


\begin{thebibliography}{72}


\ifx \showCODEN    \undefined \def \showCODEN     #1{\unskip}     \fi
\ifx \showDOI      \undefined \def \showDOI       #1{#1}\fi
\ifx \showISBNx    \undefined \def \showISBNx     #1{\unskip}     \fi
\ifx \showISBNxiii \undefined \def \showISBNxiii  #1{\unskip}     \fi
\ifx \showISSN     \undefined \def \showISSN      #1{\unskip}     \fi
\ifx \showLCCN     \undefined \def \showLCCN      #1{\unskip}     \fi
\ifx \shownote     \undefined \def \shownote      #1{#1}          \fi
\ifx \showarticletitle \undefined \def \showarticletitle #1{#1}   \fi
\ifx \showURL      \undefined \def \showURL       {\relax}        \fi
\providecommand\bibfield[2]{#2}
\providecommand\bibinfo[2]{#2}
\providecommand\natexlab[1]{#1}
\providecommand\showeprint[2][]{arXiv:#2}

\bibitem[\protect\citeauthoryear{??}{fil}{2017}]%
        {filecoin-whitepaper}
 \bibinfo{year}{2017}\natexlab{}.
\newblock \bibinfo{booktitle}{\emph{Filecoin: A Decentralized Storage
  Network}}.
\newblock \bibinfo{type}{{T}echnical {R}eport}. \bibinfo{institution}{Protocol
  Labs}.
\newblock


\bibitem[\protect\citeauthoryear{??}{nex}{2020}]%
        {nextcloud_files_external_ipfs}
 \bibinfo{year}{2020}\natexlab{}.
\newblock \bibinfo{booktitle}{\emph{{IPFS for Nextcloud}}}.
\newblock
\urldef\tempurl%
\url{https://apps.nextcloud.com/apps/files_external_ipfs}
\showURL{%
\tempurl}


\bibitem[\protect\citeauthoryear{??}{dns}{2021}]%
        {dnslink}
 \bibinfo{year}{2021}\natexlab{}.
\newblock \bibinfo{booktitle}{\emph{{DNSLink Standard}}}.
\newblock
\urldef\tempurl%
\url{https://www.dnslink.io/}
\showURL{%
\tempurl}


\bibitem[\protect\citeauthoryear{??}{mul}{2021}]%
        {multiformats}
 \bibinfo{year}{2021}\natexlab{}.
\newblock \bibinfo{booktitle}{\emph{{Multiformats -- Self-describing values for
  Future-proofing}}}.
\newblock
\urldef\tempurl%
\url{https://multiformats.io/}
\showURL{%
\tempurl}


\bibitem[\protect\citeauthoryear{??}{ipf}{2022}]%
        {ipfsecosystem}
 \bibinfo{year}{2022}\natexlab{}.
\newblock \bibinfo{title}{IPFS Ecosystem directory}.
\newblock \bibinfo{howpublished}{https://ecosystem.ipfs.io/}.
\newblock


\bibitem[\protect\citeauthoryear{??}{udg}{2022}]%
        {udgerdb}
 \bibinfo{year}{2022}\natexlab{}.
\newblock \bibinfo{title}{Udger Data v3 -- 20220606-01}.
\newblock
\newblock
\urldef\tempurl%
\url{https://udger.com/}
\showURL{%
Retrieved 02 June 2022 from \tempurl}


\bibitem[\protect\citeauthoryear{Abdullah~Lajam and Ahmed~Helmy}{Abdullah~Lajam
  and Ahmed~Helmy}{2021}]%
        {LajamH21}
\bibfield{author}{\bibinfo{person}{Omar Abdullah~Lajam} {and}
  \bibinfo{person}{Tarek Ahmed~Helmy}.} \bibinfo{year}{2021}\natexlab{}.
\newblock \showarticletitle{Performance Evaluation of IPFS in Private
  Networks}. In \bibinfo{booktitle}{\emph{2021 4th International Conference on
  Data Storage and Data Engineering}} (Barcelona, Spain)
  \emph{(\bibinfo{series}{DSDE '21})}. \bibinfo{publisher}{Association for
  Computing Machinery}, \bibinfo{address}{New York, NY, USA},
  \bibinfo{pages}{77–84}.
\newblock
\showISBNx{9781450389303}
\urldef\tempurl%
\url{https://doi.org/10.1145/3456146.3456159}
\showDOI{\tempurl}


\bibitem[\protect\citeauthoryear{Bommelaer~de Leusse and Gahnberg}{Bommelaer~de
  Leusse and Gahnberg}{2019}]%
        {bommelaer2019global}
\bibfield{author}{\bibinfo{person}{C Bommelaer~de Leusse} {and}
  \bibinfo{person}{Carl Gahnberg}.} \bibinfo{year}{2019}\natexlab{}.
\newblock \showarticletitle{The Global Internet Report: Consolidation in the
  Internet Economy}.
\newblock \bibinfo{journal}{\emph{Internet Society}} (\bibinfo{year}{2019}).
\newblock


\bibitem[\protect\citeauthoryear{B{\"o}ttger, Antichi, Fernandes, di~Lallo,
  Bruyere, Uhlig, and Castro}{B{\"o}ttger et~al\mbox{.}}{2018}]%
        {bottger2018elusive}
\bibfield{author}{\bibinfo{person}{Timm B{\"o}ttger}, \bibinfo{person}{Gianni
  Antichi}, \bibinfo{person}{Eder~L Fernandes}, \bibinfo{person}{Roberto di
  Lallo}, \bibinfo{person}{Marc Bruyere}, \bibinfo{person}{Steve Uhlig}, {and}
  \bibinfo{person}{Ignacio Castro}.} \bibinfo{year}{2018}\natexlab{}.
\newblock \showarticletitle{The elusive internet flattening: 10 years of IXP
  growth}.
\newblock \bibinfo{journal}{\emph{RIPE 78}} (\bibinfo{year}{2018}).
\newblock


\bibitem[\protect\citeauthoryear{Carzaniga, Rutherford, and Wolf}{Carzaniga
  et~al\mbox{.}}{2004}]%
        {carzaniga2004routing}
\bibfield{author}{\bibinfo{person}{Antonio Carzaniga},
  \bibinfo{person}{Matthew~J Rutherford}, {and} \bibinfo{person}{Alexander~L
  Wolf}.} \bibinfo{year}{2004}\natexlab{}.
\newblock \showarticletitle{A routing scheme for content-based networking}. In
  \bibinfo{booktitle}{\emph{IEEE INFOCOM 2004}}, Vol.~\bibinfo{volume}{2}.
  IEEE, \bibinfo{pages}{918--928}.
\newblock


\bibitem[\protect\citeauthoryear{Castro, Stanojevic, and Gorinsky}{Castro
  et~al\mbox{.}}{2013}]%
        {castro2013using}
\bibfield{author}{\bibinfo{person}{Ignacio Castro}, \bibinfo{person}{Rade
  Stanojevic}, {and} \bibinfo{person}{Sergey Gorinsky}.}
  \bibinfo{year}{2013}\natexlab{}.
\newblock \showarticletitle{Using Tuangou to reduce IP transit costs}.
\newblock \bibinfo{journal}{\emph{IEEE/ACM Transactions on Networking}}
  \bibinfo{volume}{22}, \bibinfo{number}{5} (\bibinfo{year}{2013}),
  \bibinfo{pages}{1415--1428}.
\newblock


\bibitem[\protect\citeauthoryear{Chai, Wang, Psaras, Pavlou, Wang, Garcia~de
  Blas, Ramon-Salguero, Liang, Spirou, Beben, and Hadjioannou}{Chai
  et~al\mbox{.}}{2011}]%
        {5723808}
\bibfield{author}{\bibinfo{person}{Wei~Koong Chai}, \bibinfo{person}{Ning
  Wang}, \bibinfo{person}{Ioannis Psaras}, \bibinfo{person}{George Pavlou},
  \bibinfo{person}{Chaojiong Wang}, \bibinfo{person}{Gerardo Garcia~de Blas},
  \bibinfo{person}{Francisco~Javier Ramon-Salguero}, \bibinfo{person}{Lei
  Liang}, \bibinfo{person}{Spiros Spirou}, \bibinfo{person}{Andrzej Beben},
  {and} \bibinfo{person}{Eleftheria Hadjioannou}.}
  \bibinfo{year}{2011}\natexlab{}.
\newblock \showarticletitle{Curling: Content-ubiquitous resolution and delivery
  infrastructure for next-generation services}.
\newblock \bibinfo{journal}{\emph{IEEE Communications Magazine}}
  \bibinfo{volume}{49}, \bibinfo{number}{3} (\bibinfo{year}{2011}),
  \bibinfo{pages}{112--120}.
\newblock
\urldef\tempurl%
\url{https://doi.org/10.1109/MCOM.2011.5723808}
\showDOI{\tempurl}


\bibitem[\protect\citeauthoryear{Clarke, Sandberg, Wiley, and Hong}{Clarke
  et~al\mbox{.}}{2001}]%
        {clarke2001freenet}
\bibfield{author}{\bibinfo{person}{Ian Clarke}, \bibinfo{person}{Oskar
  Sandberg}, \bibinfo{person}{Brandon Wiley}, {and} \bibinfo{person}{Theodore~W
  Hong}.} \bibinfo{year}{2001}\natexlab{}.
\newblock \showarticletitle{Freenet: A distributed anonymous information
  storage and retrieval system}. In \bibinfo{booktitle}{\emph{Designing privacy
  enhancing technologies}}. Springer.
\newblock


\bibitem[\protect\citeauthoryear{Clay}{Clay}{2013}]%
        {amazon_loss}
\bibfield{author}{\bibinfo{person}{Kelly Clay}.}
  \bibinfo{year}{2013}\natexlab{}.
\newblock \bibinfo{booktitle}{\emph{{Amazon.com Goes Down, Loses \$66,240 Per
  Minute}}}.
\newblock
\urldef\tempurl%
\url{https://web.archive.org/web/20210307232341/https://www.forbes.com/sites/kellyclay/2013/08/19/amazon-com-goes-down-loses-66240-per-minute/}
\showURL{%
\tempurl}


\bibitem[\protect\citeauthoryear{Cohen}{Cohen}{2003}]%
        {cohen2003incentives}
\bibfield{author}{\bibinfo{person}{Bram Cohen}.}
  \bibinfo{year}{2003}\natexlab{}.
\newblock \showarticletitle{Incentives build robustness in BitTorrent}. In
  \bibinfo{booktitle}{\emph{Workshop on Economics of Peer-to-Peer systems}},
  Vol.~\bibinfo{volume}{6}. Berkeley, CA, USA, \bibinfo{pages}{68--72}.
\newblock


\bibitem[\protect\citeauthoryear{Cohen}{Cohen}{2008}]%
        {bittorrent}
\bibfield{author}{\bibinfo{person}{Bram Cohen}.}
  \bibinfo{year}{2008}\natexlab{}.
\newblock \bibinfo{title}{The BitTorrent Protocol Specification v2}.
\newblock
\newblock
\urldef\tempurl%
\url{https://www.bittorrent.org/beps/bep_0052.html}
\showURL{%
Retrieved 18 May 2022 from \tempurl}


\bibitem[\protect\citeauthoryear{Coldewey}{Coldewey}{2020}]%
        {cloudFlareDNS}
\bibfield{author}{\bibinfo{person}{Devin Coldewey}.}
  \bibinfo{year}{2020}\natexlab{}.
\newblock \bibinfo{title}{Cloudflare DNS goes down, taking a large piece of the
  internet with it}.
\newblock
\newblock
\urldef\tempurl%
\url{https://techcrunch.com/2020/07/17/cloudflare-dns-goes-down-taking-a-large-piece-of-the-internet-with-it/}
\showURL{%
Retrieved 18 May 2022 from \tempurl}


\bibitem[\protect\citeauthoryear{Crosby and Wallach}{Crosby and
  Wallach}{2007}]%
        {crosby2007analysis}
\bibfield{author}{\bibinfo{person}{Scott~A Crosby} {and} \bibinfo{person}{Dan~S
  Wallach}.} \bibinfo{year}{2007}\natexlab{}.
\newblock \bibinfo{booktitle}{\emph{An analysis of bittorrent’s two
  kademlia-based dhts}}.
\newblock \bibinfo{type}{{T}echnical {R}eport}. \bibinfo{institution}{Rice
  Technical Report}.
\newblock


\bibitem[\protect\citeauthoryear{Dannewitz, Kutscher, Ohlman, Farrell, Ahlgren,
  and Karl}{Dannewitz et~al\mbox{.}}{2013}]%
        {10.1016/j.comcom.2013.01.009}
\bibfield{author}{\bibinfo{person}{Christian Dannewitz}, \bibinfo{person}{Dirk
  Kutscher}, \bibinfo{person}{B\"{o}Rje Ohlman}, \bibinfo{person}{Stephen
  Farrell}, \bibinfo{person}{Bengt Ahlgren}, {and} \bibinfo{person}{Holger
  Karl}.} \bibinfo{year}{2013}\natexlab{}.
\newblock \showarticletitle{Network of Information (NetInf) - An
  Information-Centric Networking Architecture}.
\newblock \bibinfo{journal}{\emph{Comput. Commun.}} \bibinfo{volume}{36},
  \bibinfo{number}{7} (\bibinfo{date}{apr} \bibinfo{year}{2013}),
  \bibinfo{pages}{721–735}.
\newblock
\showISSN{0140-3664}
\urldef\tempurl%
\url{https://doi.org/10.1016/j.comcom.2013.01.009}
\showDOI{\tempurl}


\bibitem[\protect\citeauthoryear{de~la Rocha, Dias, and Psaras}{de~la Rocha
  et~al\mbox{.}}{2021}]%
        {RochaDP21}
\bibfield{author}{\bibinfo{person}{Alfonso de~la Rocha}, \bibinfo{person}{David
  Dias}, {and} \bibinfo{person}{Yiannis Psaras}.}
  \bibinfo{year}{2021}\natexlab{}.
\newblock \showarticletitle{Accelerating {{Content Routing}} with {{Bitswap}}:
  {{A}} Multi-Path File Transfer Protocol in {{IPFS}} and {{Filecoin}}}.
\newblock  (\bibinfo{year}{2021}).
\newblock


\bibitem[\protect\citeauthoryear{Dias, Johnson, and Benet}{Dias
  et~al\mbox{.}}{2020}]%
        {bitswap}
\bibfield{author}{\bibinfo{person}{David Dias}, \bibinfo{person}{Jeromy
  Johnson}, {and} \bibinfo{person}{Juan Benet}.}
  \bibinfo{year}{2020}\natexlab{}.
\newblock \bibinfo{title}{Bitswap -- Protocol Specification}.
\newblock
\newblock
\urldef\tempurl%
\url{https://github.com/ipfs/specs/blob/master/BITSWAP.md}
\showURL{%
Retrieved 01 June 2022 from \tempurl}


\bibitem[\protect\citeauthoryear{Falkner, Piatek, John, Krishnamurthy, and
  Anderson}{Falkner et~al\mbox{.}}{2007}]%
        {10.1145/1298306.1298325}
\bibfield{author}{\bibinfo{person}{Jarret Falkner}, \bibinfo{person}{Michael
  Piatek}, \bibinfo{person}{John~P. John}, \bibinfo{person}{Arvind
  Krishnamurthy}, {and} \bibinfo{person}{Thomas Anderson}.}
  \bibinfo{year}{2007}\natexlab{}.
\newblock \showarticletitle{Profiling a Million User Dht}. In
  \bibinfo{booktitle}{\emph{Proceedings of the 7th ACM SIGCOMM Conference on
  Internet Measurement}} (San Diego, California, USA)
  \emph{(\bibinfo{series}{IMC '07})}. \bibinfo{publisher}{Association for
  Computing Machinery}, \bibinfo{address}{New York, NY, USA},
  \bibinfo{pages}{129–134}.
\newblock
\showISBNx{9781595939081}
\urldef\tempurl%
\url{https://doi.org/10.1145/1298306.1298325}
\showDOI{\tempurl}


\bibitem[\protect\citeauthoryear{Fanou, Tyson, Francois, and
  Sathiaseelan}{Fanou et~al\mbox{.}}{2016}]%
        {fanou2016pushing}
\bibfield{author}{\bibinfo{person}{Rod{\'e}rick Fanou}, \bibinfo{person}{Gareth
  Tyson}, \bibinfo{person}{Pierre Francois}, {and} \bibinfo{person}{Arjuna
  Sathiaseelan}.} \bibinfo{year}{2016}\natexlab{}.
\newblock \showarticletitle{Pushing the frontier: Exploring the african web
  ecosystem}. In \bibinfo{booktitle}{\emph{Proceedings of the 25th
  International Conference on World Wide Web}}.
  \bibinfo{publisher}{International World Wide Web Conferences Steering
  Committee}, \bibinfo{pages}{435--445}.
\newblock
\urldef\tempurl%
\url{https://doi.org/10.1145/2872427.2882997}
\showDOI{\tempurl}


\bibitem[\protect\citeauthoryear{Gnutella}{Gnutella}{2009}]%
        {gnutella}
\bibfield{author}{\bibinfo{person}{Gnutella}.} \bibinfo{year}{2009}\natexlab{}.
\newblock \bibinfo{title}{Gnutella Protocol Specification}.
\newblock
\newblock
\urldef\tempurl%
\url{https://web.archive.org/web/20090331221153/http://wiki.limewire.org/index.php?title=GDF}
\showURL{%
Retrieved 18 May 2022 from \tempurl}


\bibitem[\protect\citeauthoryear{Graffi, Gross, Stingl, Hartung, Kovacevic, and
  Steinmetz}{Graffi et~al\mbox{.}}{2011}]%
        {graffi2011lifesocial}
\bibfield{author}{\bibinfo{person}{Kalman Graffi}, \bibinfo{person}{Christian
  Gross}, \bibinfo{person}{Dominik Stingl}, \bibinfo{person}{Daniel Hartung},
  \bibinfo{person}{Aleksandra Kovacevic}, {and} \bibinfo{person}{Ralf
  Steinmetz}.} \bibinfo{year}{2011}\natexlab{}.
\newblock \showarticletitle{{LifeSocial. KOM: A secure and P2P-based solution
  for online social networks}}. In \bibinfo{booktitle}{\emph{{CCNC}}}.
\newblock


\bibitem[\protect\citeauthoryear{Guidi, Conti, Passarella, and Ricci}{Guidi
  et~al\mbox{.}}{2018}]%
        {guidi2018managing}
\bibfield{author}{\bibinfo{person}{Barbara Guidi}, \bibinfo{person}{Marco
  Conti}, \bibinfo{person}{Andrea Passarella}, {and} \bibinfo{person}{Laura
  Ricci}.} \bibinfo{year}{2018}\natexlab{}.
\newblock \showarticletitle{{Managing social contents in Decentralized Online
  Social Networks: A survey}}.
\newblock \bibinfo{journal}{\emph{Online Social Networks and Media}}
  \bibinfo{volume}{7} (\bibinfo{year}{2018}).
\newblock


\bibitem[\protect\citeauthoryear{Hassan, Raman, Castro, Zia, De~Cristofaro,
  Sastry, and Tyson}{Hassan et~al\mbox{.}}{2021}]%
        {hassan2021exploring}
\bibfield{author}{\bibinfo{person}{Anaobi~Ishaku Hassan},
  \bibinfo{person}{Aravindh Raman}, \bibinfo{person}{Ignacio Castro},
  \bibinfo{person}{Haris~Bin Zia}, \bibinfo{person}{Emiliano De~Cristofaro},
  \bibinfo{person}{Nishanth Sastry}, {and} \bibinfo{person}{Gareth Tyson}.}
  \bibinfo{year}{2021}\natexlab{}.
\newblock \showarticletitle{Exploring content moderation in the decentralised
  web: The pleroma case}. In \bibinfo{booktitle}{\emph{Proceedings of the 17th
  International Conference on emerging Networking EXperiments and
  Technologies}}. \bibinfo{pages}{328--335}.
\newblock


\bibitem[\protect\citeauthoryear{Henningsen, Florian, Rust, and
  Scheuermann}{Henningsen et~al\mbox{.}}{2020}]%
        {HenningsenFRS20}
\bibfield{author}{\bibinfo{person}{Sebastian~A. Henningsen},
  \bibinfo{person}{Martin Florian}, \bibinfo{person}{Sebastian Rust}, {and}
  \bibinfo{person}{Bj{\"o}rn Scheuermann}.} \bibinfo{year}{2020}\natexlab{}.
\newblock \showarticletitle{Mapping the Interplanetary Filesystem}.
\newblock \bibinfo{journal}{\emph{2020 IFIP Networking Conference
  (Networking)}} (\bibinfo{year}{2020}), \bibinfo{pages}{289--297}.
\newblock


\bibitem[\protect\citeauthoryear{Holz, Hiller, Amann, Razaghpanah, Jost,
  Vallina-Rodriguez, and Hohlfeld}{Holz et~al\mbox{.}}{2020}]%
        {holz2020tracking}
\bibfield{author}{\bibinfo{person}{Ralph Holz}, \bibinfo{person}{Jens Hiller},
  \bibinfo{person}{Johanna Amann}, \bibinfo{person}{Abbas Razaghpanah},
  \bibinfo{person}{Thomas Jost}, \bibinfo{person}{Narseo Vallina-Rodriguez},
  {and} \bibinfo{person}{Oliver Hohlfeld}.} \bibinfo{year}{2020}\natexlab{}.
\newblock \showarticletitle{Tracking the deployment of TLS 1.3 on the Web: A
  story of experimentation and centralization}.
\newblock \bibinfo{journal}{\emph{ACM SIGCOMM Computer Communication Review}}
  \bibinfo{volume}{50}, \bibinfo{number}{3} (\bibinfo{year}{2020}),
  \bibinfo{pages}{3--15}.
\newblock


\bibitem[\protect\citeauthoryear{Isdal, Piatek, Krishnamurthy, and
  Anderson}{Isdal et~al\mbox{.}}{2010}]%
        {10.1145/1851182.1851198}
\bibfield{author}{\bibinfo{person}{Tomas Isdal}, \bibinfo{person}{Michael
  Piatek}, \bibinfo{person}{Arvind Krishnamurthy}, {and}
  \bibinfo{person}{Thomas Anderson}.} \bibinfo{year}{2010}\natexlab{}.
\newblock \showarticletitle{Privacy-Preserving P2P Data Sharing with OneSwarm}.
  In \bibinfo{booktitle}{\emph{Proceedings of the ACM SIGCOMM 2010 Conference}}
  (New Delhi, India) \emph{(\bibinfo{series}{SIGCOMM '10})}.
  \bibinfo{publisher}{Association for Computing Machinery},
  \bibinfo{address}{New York, NY, USA}, \bibinfo{pages}{111–122}.
\newblock
\showISBNx{9781450302012}
\urldef\tempurl%
\url{https://doi.org/10.1145/1851182.1851198}
\showDOI{\tempurl}


\bibitem[\protect\citeauthoryear{Jacobson, Smetters, Thornton, Plass, Briggs,
  and Braynard}{Jacobson et~al\mbox{.}}{2009}]%
        {10.1145/1658939.1658941}
\bibfield{author}{\bibinfo{person}{Van Jacobson}, \bibinfo{person}{Diana~K.
  Smetters}, \bibinfo{person}{James~D. Thornton}, \bibinfo{person}{Michael~F.
  Plass}, \bibinfo{person}{Nicholas~H. Briggs}, {and}
  \bibinfo{person}{Rebecca~L. Braynard}.} \bibinfo{year}{2009}\natexlab{}.
\newblock \showarticletitle{Networking Named Content}. In
  \bibinfo{booktitle}{\emph{Proceedings of the 5th International Conference on
  Emerging Networking Experiments and Technologies}} (Rome, Italy)
  \emph{(\bibinfo{series}{CoNEXT '09})}. \bibinfo{publisher}{Association for
  Computing Machinery}, \bibinfo{address}{New York, NY, USA},
  \bibinfo{pages}{1–12}.
\newblock
\showISBNx{9781605586366}
\urldef\tempurl%
\url{https://doi.org/10.1145/1658939.1658941}
\showDOI{\tempurl}


\bibitem[\protect\citeauthoryear{Kaashoek and Karger}{Kaashoek and
  Karger}{2003}]%
        {kaashoek2003koorde}
\bibfield{author}{\bibinfo{person}{M~Frans Kaashoek} {and}
  \bibinfo{person}{David~R Karger}.} \bibinfo{year}{2003}\natexlab{}.
\newblock \showarticletitle{Koorde: A simple degree-optimal distributed hash
  table}. In \bibinfo{booktitle}{\emph{International Workshop on Peer-to-Peer
  Systems}}. Springer, \bibinfo{pages}{98--107}.
\newblock


\bibitem[\protect\citeauthoryear{Kaune, Pussep, Leng, Kovacevic, Tyson, and
  Steinmetz}{Kaune et~al\mbox{.}}{2009}]%
        {kaune2009modelling}
\bibfield{author}{\bibinfo{person}{Sebastian Kaune},
  \bibinfo{person}{Konstantin Pussep}, \bibinfo{person}{Christof Leng},
  \bibinfo{person}{Aleksandra Kovacevic}, \bibinfo{person}{Gareth Tyson}, {and}
  \bibinfo{person}{Ralf Steinmetz}.} \bibinfo{year}{2009}\natexlab{}.
\newblock \showarticletitle{Modelling the internet delay space based on
  geographical locations}. In \bibinfo{booktitle}{\emph{2009 17th Euromicro
  International Conference on Parallel, Distributed and Network-based
  Processing}}. IEEE, \bibinfo{pages}{301--310}.
\newblock


\bibitem[\protect\citeauthoryear{Khare, Karan, McQuistin, Perkins, Tyson,
  Purver, Healey, and Castro}{Khare et~al\mbox{.}}{2022}]%
        {khare2022web}
\bibfield{author}{\bibinfo{person}{Prashant Khare}, \bibinfo{person}{Mladen
  Karan}, \bibinfo{person}{Stephen McQuistin}, \bibinfo{person}{Colin Perkins},
  \bibinfo{person}{Gareth Tyson}, \bibinfo{person}{Matthew Purver},
  \bibinfo{person}{Patrick Healey}, {and} \bibinfo{person}{Ignacio Castro}.}
  \bibinfo{year}{2022}\natexlab{}.
\newblock \showarticletitle{The Web We Weave: Untangling the Social Graph of
  the IETF}. In \bibinfo{booktitle}{\emph{Proceedings of the International AAAI
  Conference on Web and Social Media}}, Vol.~\bibinfo{volume}{16}.
  \bibinfo{pages}{500--511}.
\newblock


\bibitem[\protect\citeauthoryear{Koll, Lechler, and Fu}{Koll
  et~al\mbox{.}}{2017}]%
        {koll2017socialgate}
\bibfield{author}{\bibinfo{person}{David Koll}, \bibinfo{person}{Dieter
  Lechler}, {and} \bibinfo{person}{Xiaoming Fu}.}
  \bibinfo{year}{2017}\natexlab{}.
\newblock \showarticletitle{{SocialGate: Managing large-scale social data on
  home gateways}}. In \bibinfo{booktitle}{\emph{{IEEE ICNP}}}.
\newblock


\bibitem[\protect\citeauthoryear{Labs}{Labs}{2021a}]%
        {dcutr}
\bibfield{author}{\bibinfo{person}{Protocol Labs}.}
  \bibinfo{year}{2021}\natexlab{a}.
\newblock \bibinfo{title}{Direct Connection Upgrade through Relay}.
\newblock
\newblock
\urldef\tempurl%
\url{https://github.com/libp2p/specs/blob/master/relay/DCUtR.md}
\showURL{%
Retrieved 01 June 2022 from \tempurl}


\bibitem[\protect\citeauthoryear{Labs}{Labs}{2021b}]%
        {merkleDAG}
\bibfield{author}{\bibinfo{person}{Protocol Labs}.}
  \bibinfo{year}{2021}\natexlab{b}.
\newblock \bibinfo{title}{Merkle Directed Acyclic Graphs (DAGs)}.
\newblock
\newblock
\urldef\tempurl%
\url{https://docs.ipfs.io/concepts/merkle-dag/}
\showURL{%
Retrieved 18 May 2022 from \tempurl}


\bibitem[\protect\citeauthoryear{Labs}{Labs}{2022}]%
        {autonat}
\bibfield{author}{\bibinfo{person}{Protocol Labs}.}
  \bibinfo{year}{2022}\natexlab{}.
\newblock \bibinfo{title}{AutoNAT}.
\newblock
\newblock
\urldef\tempurl%
\url{https://github.com/libp2p/specs/blob/master/autonat/README.md}
\showURL{%
Retrieved 01 June 2022 from \tempurl}


\bibitem[\protect\citeauthoryear{{Le Pochat}, {Van Goethem}, Tajalizadehkhoob,
  Korczy\'{n}ski, and Joosen}{{Le Pochat} et~al\mbox{.}}{2019}]%
        {LePochat2019tranco}
\bibfield{author}{\bibinfo{person}{Victor {Le Pochat}}, \bibinfo{person}{Tom
  {Van Goethem}}, \bibinfo{person}{Samaneh Tajalizadehkhoob},
  \bibinfo{person}{Maciej Korczy\'{n}ski}, {and} \bibinfo{person}{Wouter
  Joosen}.} \bibinfo{year}{2019}\natexlab{}.
\newblock \showarticletitle{Tranco: A Research-Oriented Top Sites Ranking
  Hardened Against Manipulation}. In \bibinfo{booktitle}{\emph{Proceedings of
  the 26th Annual Network and Distributed System Security Symposium}}
  \emph{(\bibinfo{series}{NDSS 2019})}.
\newblock
\urldef\tempurl%
\url{https://doi.org/10.14722/ndss.2019.23386}
\showDOI{\tempurl}


\bibitem[\protect\citeauthoryear{Li, Muthusamy, and Jacobsen}{Li
  et~al\mbox{.}}{2008}]%
        {li2008adaptive}
\bibfield{author}{\bibinfo{person}{Guoli Li}, \bibinfo{person}{Vinod
  Muthusamy}, {and} \bibinfo{person}{Hans-Arno Jacobsen}.}
  \bibinfo{year}{2008}\natexlab{}.
\newblock \showarticletitle{Adaptive content-based routing in general overlay
  topologies}. In \bibinfo{booktitle}{\emph{ACM/IFIP/USENIX International
  Conference on Distributed Systems Platforms and Open Distributed
  Processing}}. Springer, \bibinfo{pages}{1--21}.
\newblock


\bibitem[\protect\citeauthoryear{Mager, Biersack, and Michiardi}{Mager
  et~al\mbox{.}}{2012}]%
        {mager2012measurement}
\bibfield{author}{\bibinfo{person}{Thomas Mager}, \bibinfo{person}{Ernst
  Biersack}, {and} \bibinfo{person}{Pietro Michiardi}.}
  \bibinfo{year}{2012}\natexlab{}.
\newblock \showarticletitle{A measurement study of the Wuala on-line storage
  service}. In \bibinfo{booktitle}{\emph{2012 IEEE 12th International
  Conference on Peer-to-Peer Computing (P2P)}}. IEEE.
\newblock


\bibitem[\protect\citeauthoryear{Mansour, Sambra, Hawke, Zereba, Capadisli,
  Ghanem, Aboulnaga, and Berners-Lee}{Mansour et~al\mbox{.}}{2016}]%
        {mansour2016demonstration}
\bibfield{author}{\bibinfo{person}{Essam Mansour}, \bibinfo{person}{Andrei~Vlad
  Sambra}, \bibinfo{person}{Sandro Hawke}, \bibinfo{person}{Maged Zereba},
  \bibinfo{person}{Sarven Capadisli}, \bibinfo{person}{Abdurrahman Ghanem},
  \bibinfo{person}{Ashraf Aboulnaga}, {and} \bibinfo{person}{Tim Berners-Lee}.}
  \bibinfo{year}{2016}\natexlab{}.
\newblock \showarticletitle{{A demonstration of the solid platform for social
  web applications}}. In \bibinfo{booktitle}{\emph{{WWW}}}.
\newblock


\bibitem[\protect\citeauthoryear{Mathews}{Mathews}{2021}]%
        {awsOutage}
\bibfield{author}{\bibinfo{person}{Eva Mathews}.}
  \bibinfo{year}{2021}\natexlab{}.
\newblock \bibinfo{title}{Amazon cloud outage hits major websites, streaming
  apps}.
\newblock
\newblock
\urldef\tempurl%
\url{https://www.reuters.com/article/amazon-com-outages-idCAKBN2IM1U0}
\showURL{%
Retrieved 18 May 2022 from \tempurl}


\bibitem[\protect\citeauthoryear{Maymounkov and Mazieres}{Maymounkov and
  Mazieres}{2002}]%
        {maymounkov2002kademlia}
\bibfield{author}{\bibinfo{person}{Petar Maymounkov} {and}
  \bibinfo{person}{David Mazieres}.} \bibinfo{year}{2002}\natexlab{}.
\newblock \showarticletitle{Kademlia: A peer-to-peer information system based
  on the xor metric}. In \bibinfo{booktitle}{\emph{International Workshop on
  Peer-to-Peer Systems}}. Springer, \bibinfo{pages}{53--65}.
\newblock


\bibitem[\protect\citeauthoryear{McQuistin, Karan, Khare, Perkins, Tyson,
  Purver, Healey, Iqbal, Qadir, and Castro}{McQuistin et~al\mbox{.}}{2021}]%
        {mcquistin2021characterising}
\bibfield{author}{\bibinfo{person}{Stephen McQuistin}, \bibinfo{person}{Mladen
  Karan}, \bibinfo{person}{Prashant Khare}, \bibinfo{person}{Colin Perkins},
  \bibinfo{person}{Gareth Tyson}, \bibinfo{person}{Matthew Purver},
  \bibinfo{person}{Patrick Healey}, \bibinfo{person}{Waleed Iqbal},
  \bibinfo{person}{Junaid Qadir}, {and} \bibinfo{person}{Ignacio Castro}.}
  \bibinfo{year}{2021}\natexlab{}.
\newblock \showarticletitle{Characterising the IETF through the lens of RFC
  deployment}. In \bibinfo{booktitle}{\emph{Proceedings of the 21st ACM
  Internet Measurement Conference}}. \bibinfo{pages}{137--149}.
\newblock


\bibitem[\protect\citeauthoryear{Parrillo and Tschudin}{Parrillo and
  Tschudin}{2021}]%
        {9472772}
\bibfield{author}{\bibinfo{person}{Fabrizio Parrillo} {and}
  \bibinfo{person}{Christian Tschudin}.} \bibinfo{year}{2021}\natexlab{}.
\newblock \showarticletitle{Solid over the Interplanetary File System}. In
  \bibinfo{booktitle}{\emph{2021 IFIP Networking Conference (IFIP
  Networking)}}.
\newblock


\bibitem[\protect\citeauthoryear{Perera, Wakenshaw, Baarslag, Haddadi, Bandara,
  Mortier, Crabtree, Ng, McAuley, and Crowcroft}{Perera et~al\mbox{.}}{2017}]%
        {perera2017valorising}
\bibfield{author}{\bibinfo{person}{Charith Perera}, \bibinfo{person}{Susan~YL
  Wakenshaw}, \bibinfo{person}{Tim Baarslag}, \bibinfo{person}{Hamed Haddadi},
  \bibinfo{person}{Arosha~K Bandara}, \bibinfo{person}{Richard Mortier},
  \bibinfo{person}{Andy Crabtree}, \bibinfo{person}{Irene~CL Ng},
  \bibinfo{person}{Derek McAuley}, {and} \bibinfo{person}{Jon Crowcroft}.}
  \bibinfo{year}{2017}\natexlab{}.
\newblock \showarticletitle{{Valorising the IoT databox: creating value for
  everyone}}.
\newblock \bibinfo{journal}{\emph{Transactions on Emerging Telecommunications
  Technologies}} \bibinfo{volume}{28}, \bibinfo{number}{1}
  (\bibinfo{year}{2017}).
\newblock


\bibitem[\protect\citeauthoryear{Piatek, Isdal, Anderson, Krishnamurthy, and
  Venkataramani}{Piatek et~al\mbox{.}}{2007}]%
        {202577}
\bibfield{author}{\bibinfo{person}{Michael Piatek}, \bibinfo{person}{Tomas
  Isdal}, \bibinfo{person}{Thomas Anderson}, \bibinfo{person}{Arvind
  Krishnamurthy}, {and} \bibinfo{person}{Arun Venkataramani}.}
  \bibinfo{year}{2007}\natexlab{}.
\newblock \showarticletitle{Do Incentives Build Robustness in {BitTorrent}?}.
  In \bibinfo{booktitle}{\emph{4th USENIX Symposium on Networked Systems Design
  \& Implementation (NSDI 07)}}. \bibinfo{publisher}{USENIX Association},
  \bibinfo{address}{Cambridge, MA}.
\newblock
\urldef\tempurl%
\url{https://www.usenix.org/conference/nsdi-07/do-incentives-build-robustness-bittorrent}
\showURL{%
\tempurl}


\bibitem[\protect\citeauthoryear{Raman, Joglekar, Cristofaro, Sastry, and
  Tyson}{Raman et~al\mbox{.}}{2019}]%
        {raman2019challenges}
\bibfield{author}{\bibinfo{person}{Aravindh Raman}, \bibinfo{person}{Sagar
  Joglekar}, \bibinfo{person}{Emiliano~De Cristofaro},
  \bibinfo{person}{Nishanth Sastry}, {and} \bibinfo{person}{Gareth Tyson}.}
  \bibinfo{year}{2019}\natexlab{}.
\newblock \showarticletitle{Challenges in the decentralised web: The mastodon
  case}. In \bibinfo{booktitle}{\emph{Proceedings of the Internet Measurement
  Conference}}. \bibinfo{pages}{217--229}.
\newblock


\bibitem[\protect\citeauthoryear{Rank}{Rank}{2022}]%
        {caida_asrank}
\bibfield{author}{\bibinfo{person}{CAIDA~AS Rank}.}
  \bibinfo{year}{2022}\natexlab{}.
\newblock \bibinfo{howpublished}{{url{http://as-rank.caida.org/}}}.
\newblock


\bibitem[\protect\citeauthoryear{Ratnasamy, Francis, Handley, Karp, and
  Shenker}{Ratnasamy et~al\mbox{.}}{2001}]%
        {10.1145/964723.383072}
\bibfield{author}{\bibinfo{person}{Sylvia Ratnasamy}, \bibinfo{person}{Paul
  Francis}, \bibinfo{person}{Mark Handley}, \bibinfo{person}{Richard Karp},
  {and} \bibinfo{person}{Scott Shenker}.} \bibinfo{year}{2001}\natexlab{}.
\newblock \showarticletitle{A Scalable Content-Addressable Network}.
\newblock \bibinfo{journal}{\emph{SIGCOMM Comput. Commun. Rev.}}
  \bibinfo{volume}{31}, \bibinfo{number}{4} (\bibinfo{date}{aug}
  \bibinfo{year}{2001}), \bibinfo{pages}{161–172}.
\newblock
\showISSN{0146-4833}
\urldef\tempurl%
\url{https://doi.org/10.1145/964723.383072}
\showDOI{\tempurl}


\bibitem[\protect\citeauthoryear{Roselli, Lorch, and Anderson}{Roselli
  et~al\mbox{.}}{[n.d.]}]%
        {RoselliLA}
\bibfield{author}{\bibinfo{person}{Drew Roselli}, \bibinfo{person}{Jacob~R
  Lorch}, {and} \bibinfo{person}{Thomas~E Anderson}.}
  \bibinfo{year}{[n.d.]}\natexlab{}.
\newblock \showarticletitle{A {{Comparison}} of {{File System Workloads}}}.
\newblock  (\bibinfo{year}{[n.\,d.]}), \bibinfo{pages}{14}.
\newblock


\bibitem[\protect\citeauthoryear{Rosemain and Satter}{Rosemain and
  Satter}{2021}]%
        {ovhOutage}
\bibfield{author}{\bibinfo{person}{Mathieu Rosemain} {and}
  \bibinfo{person}{Raphael Satter}.} \bibinfo{year}{2021}\natexlab{}.
\newblock \bibinfo{title}{Millions of websites offline after fire at French
  cloud services firm}.
\newblock
\newblock
\urldef\tempurl%
\url{https://www.reuters.com/article/us-france-ovh-fire-idUSKBN2B20NU}
\showURL{%
Retrieved 18 May 2022 from \tempurl}


\bibitem[\protect\citeauthoryear{Rowstron and Druschel}{Rowstron and
  Druschel}{2001}]%
        {rowstron2001pastry}
\bibfield{author}{\bibinfo{person}{Antony Rowstron} {and}
  \bibinfo{person}{Peter Druschel}.} \bibinfo{year}{2001}\natexlab{}.
\newblock \showarticletitle{Pastry: Scalable, decentralized object location,
  and routing for large-scale peer-to-peer systems}. In
  \bibinfo{booktitle}{\emph{IFIP/ACM International Conference on Distributed
  Systems Platforms and Open Distributed Processing}}. Springer,
  \bibinfo{pages}{329--350}.
\newblock


\bibitem[\protect\citeauthoryear{Saleh and Hefeeda}{Saleh and Hefeeda}{2006}]%
        {saleh2006modeling}
\bibfield{author}{\bibinfo{person}{Osama Saleh} {and} \bibinfo{person}{Mohamed
  Hefeeda}.} \bibinfo{year}{2006}\natexlab{}.
\newblock \showarticletitle{Modeling and Caching of Peer-to-Peer Traffic}. In
  \bibinfo{booktitle}{\emph{Proceedings of the 2006 {IEEE} International
  Conference on Network Protocols}}. \bibinfo{publisher}{{IEEE}}.
\newblock
\urldef\tempurl%
\url{https://doi.org/10.1109/icnp.2006.320218}
\showDOI{\tempurl}


\bibitem[\protect\citeauthoryear{Santos and Allen}{Santos and Allen}{2021}]%
        {ipns}
\bibfield{author}{\bibinfo{person}{Vasco Santos} {and} \bibinfo{person}{Steven
  Allen}.} \bibinfo{year}{2021}\natexlab{}.
\newblock \bibinfo{booktitle}{\emph{{IPNS - Inter-Planetary Naming System}}}.
\newblock
\urldef\tempurl%
\url{https://github.com/ipfs/specs/blob/bab189ee61c316eb3d371c7270abef97641e7ed9/IPNS.md}
\showURL{%
\tempurl}


\bibitem[\protect\citeauthoryear{Saroiu, Gummadi, and Gribble}{Saroiu
  et~al\mbox{.}}{[n.d.]}]%
        {SaroiuGG}
\bibfield{author}{\bibinfo{person}{Stefan Saroiu}, \bibinfo{person}{P~Krishna
  Gummadi}, {and} \bibinfo{person}{Steven~D Gribble}.}
  \bibinfo{year}{[n.d.]}\natexlab{}.
\newblock \showarticletitle{A {{Measurement Study}} of {{Peer-to-Peer File
  Sharing Systems}}}. In \bibinfo{booktitle}{\emph{Multimedia {{Computing}} and
  {{Networking}} 2002}} (2001-12-10), Vol.~\bibinfo{volume}{4673}.
  \bibinfo{publisher}{{SPIE}}, \bibinfo{pages}{156--170}.
\newblock
\urldef\tempurl%
\url{https://doi.org/10.1117/12.449977}
\showDOI{\tempurl}


\bibitem[\protect\citeauthoryear{Schwittmann, Boelmann, Wander, and
  Weis}{Schwittmann et~al\mbox{.}}{2013}]%
        {schwittmann2013sonet}
\bibfield{author}{\bibinfo{person}{Lorenz Schwittmann},
  \bibinfo{person}{Christopher Boelmann}, \bibinfo{person}{Matthaus Wander},
  {and} \bibinfo{person}{Torben Weis}.} \bibinfo{year}{2013}\natexlab{}.
\newblock \showarticletitle{{SoNet--Privacy and Replication in Federated Online
  Social Networks}}. In \bibinfo{booktitle}{\emph{{Distributed Computing
  Systems Workshops}}}.
\newblock


\bibitem[\protect\citeauthoryear{Stoica, Morris, Karger, Kaashoek, and
  Balakrishnan}{Stoica et~al\mbox{.}}{2001}]%
        {stoica2001chord}
\bibfield{author}{\bibinfo{person}{Ion Stoica}, \bibinfo{person}{Robert
  Morris}, \bibinfo{person}{David Karger}, \bibinfo{person}{M~Frans Kaashoek},
  {and} \bibinfo{person}{Hari Balakrishnan}.} \bibinfo{year}{2001}\natexlab{}.
\newblock \showarticletitle{Chord: A scalable peer-to-peer lookup service for
  internet applications}.
\newblock \bibinfo{journal}{\emph{ACM SIGCOMM Computer Communication Review}}
  \bibinfo{volume}{31}, \bibinfo{number}{4} (\bibinfo{year}{2001}),
  \bibinfo{pages}{149--160}.
\newblock


\bibitem[\protect\citeauthoryear{Stutzbach and Rejaie}{Stutzbach and
  Rejaie}{2006a}]%
        {stutzbach2006improving}
\bibfield{author}{\bibinfo{person}{Daniel Stutzbach} {and}
  \bibinfo{person}{Reza Rejaie}.} \bibinfo{year}{2006}\natexlab{a}.
\newblock \showarticletitle{Improving lookup performance over a widely-deployed
  DHT}. In \bibinfo{booktitle}{\emph{Proceedings IEEE INFOCOM 2006. 25TH IEEE
  International Conference on Computer Communications}}. IEEE,
  \bibinfo{pages}{1--12}.
\newblock


\bibitem[\protect\citeauthoryear{Stutzbach and Rejaie}{Stutzbach and
  Rejaie}{2006b}]%
        {StutzbachR06}
\bibfield{author}{\bibinfo{person}{Daniel Stutzbach} {and}
  \bibinfo{person}{Reza Rejaie}.} \bibinfo{year}{2006}\natexlab{b}.
\newblock \showarticletitle{Understanding Churn in Peer-to-Peer Networks}. In
  \bibinfo{booktitle}{\emph{Proceedings of the 6th ACM SIGCOMM Conference on
  Internet Measurement}} (Rio de Janeiro, Brazil) \emph{(\bibinfo{series}{IMC
  '06})}. \bibinfo{publisher}{Association for Computing Machinery},
  \bibinfo{address}{New York, NY, USA}, \bibinfo{pages}{189–202}.
\newblock
\showISBNx{1595935614}
\urldef\tempurl%
\url{https://doi.org/10.1145/1177080.1177105}
\showDOI{\tempurl}


\bibitem[\protect\citeauthoryear{Taheri-Boshrooyeh, K{\"u}p{\c{c}}{\"u}, and
  {\"O}zkasap}{Taheri-Boshrooyeh et~al\mbox{.}}{2015}]%
        {taheri2015security}
\bibfield{author}{\bibinfo{person}{Sanaz Taheri-Boshrooyeh},
  \bibinfo{person}{Alptekin K{\"u}p{\c{c}}{\"u}}, {and}
  \bibinfo{person}{{\"O}znur {\"O}zkasap}.} \bibinfo{year}{2015}\natexlab{}.
\newblock \showarticletitle{{Security and privacy of distributed online social
  networks}}. In \bibinfo{booktitle}{\emph{{Distributed Computing Systems
  Workshops}}}.
\newblock


\bibitem[\protect\citeauthoryear{Tarr, Lavoie, Meyer, and Tschudin}{Tarr
  et~al\mbox{.}}{2019}]%
        {10.1145/3357150.3357396}
\bibfield{author}{\bibinfo{person}{Dominic Tarr}, \bibinfo{person}{Erick
  Lavoie}, \bibinfo{person}{Aljoscha Meyer}, {and} \bibinfo{person}{Christian
  Tschudin}.} \bibinfo{year}{2019}\natexlab{}.
\newblock \showarticletitle{Secure Scuttlebutt: An Identity-Centric Protocol
  for Subjective and Decentralized Applications}. In
  \bibinfo{booktitle}{\emph{Proceedings of the 6th ACM Conference on
  Information-Centric Networking}} (Macao, China) \emph{(\bibinfo{series}{ICN
  '19})}. \bibinfo{publisher}{Association for Computing Machinery},
  \bibinfo{address}{New York, NY, USA}, \bibinfo{pages}{1–11}.
\newblock
\showISBNx{9781450369701}
\urldef\tempurl%
\url{https://doi.org/10.1145/3357150.3357396}
\showDOI{\tempurl}


\bibitem[\protect\citeauthoryear{Trautwein}{Trautwein}{2021}]%
        {Trautwein_Nebula21}
\bibfield{author}{\bibinfo{person}{Dennis Trautwein}.}
  \bibinfo{year}{2021}\natexlab{}.
\newblock \bibinfo{booktitle}{\emph{{Nebula -- A crawler for networks based on
  the libp2p DHT implementation}}}.
\newblock
\urldef\tempurl%
\url{https://github.com/dennis-tra/nebula-crawler}
\showURL{%
\tempurl}


\bibitem[\protect\citeauthoryear{Vorick and Champine}{Vorick and
  Champine}{2014}]%
        {sia-whitepaper}
\bibfield{author}{\bibinfo{person}{David Vorick} {and} \bibinfo{person}{Luke
  Champine}.} \bibinfo{year}{2014}\natexlab{}.
\newblock \bibinfo{booktitle}{\emph{Sia: Simple Decentralized Storage}}.
\newblock \bibinfo{type}{{T}echnical {R}eport}. \bibinfo{institution}{Nebulous
  Inc.}
\newblock


\bibitem[\protect\citeauthoryear{Wang and Kangasharju}{Wang and
  Kangasharju}{2013}]%
        {6688697}
\bibfield{author}{\bibinfo{person}{Liang Wang} {and} \bibinfo{person}{J.
  Kangasharju}.} \bibinfo{year}{2013}\natexlab{}.
\newblock \showarticletitle{Measuring large-scale distributed systems: case of
  BitTorrent Mainline DHT}. In \bibinfo{booktitle}{\emph{Peer-to-Peer Computing
  (P2P), 2013 IEEE Thirteenth International Conference on}}.
  \bibinfo{pages}{1--10}.
\newblock
\urldef\tempurl%
\url{https://doi.org/10.1109/P2P.2013.6688697}
\showDOI{\tempurl}


\bibitem[\protect\citeauthoryear{Wang, Yin, and Yu}{Wang et~al\mbox{.}}{2005}]%
        {Wang2005}
\bibfield{author}{\bibinfo{person}{Xiaoyun Wang}, \bibinfo{person}{Yiqun~Lisa
  Yin}, {and} \bibinfo{person}{Hongbo Yu}.} \bibinfo{year}{2005}\natexlab{}.
\newblock \showarticletitle{Finding Collisions in the Full {SHA}-1}.
\newblock In \bibinfo{booktitle}{\emph{Advances in Cryptology {\textendash}
  {CRYPTO} 2005}}. \bibinfo{publisher}{Springer Berlin Heidelberg},
  \bibinfo{pages}{17--36}.
\newblock
\urldef\tempurl%
\url{https://doi.org/10.1007/11535218_2}
\showDOI{\tempurl}


\bibitem[\protect\citeauthoryear{{Web Storage Index}}{{Web Storage
  Index}}{2022}]%
        {storageindex}
\bibfield{author}{\bibinfo{person}{{Web Storage Index}}.}
  \bibinfo{year}{2022}\natexlab{}.
\newblock \bibinfo{howpublished}{\url{https://www.storageindex.io/}}.
\newblock


\bibitem[\protect\citeauthoryear{Williams, Diordiiev, Berman, Raybould, and
  Uemlianin}{Williams et~al\mbox{.}}{[n.d.]}]%
        {arweave-whitepaper}
\bibfield{author}{\bibinfo{person}{Sam Williams}, \bibinfo{person}{Viktor
  Diordiiev}, \bibinfo{person}{Lev Berman}, \bibinfo{person}{India Raybould},
  {and} \bibinfo{person}{Ivan Uemlianin}.} \bibinfo{year}{[n.d.]}\natexlab{}.
\newblock \bibinfo{booktitle}{\emph{Arweave: A Protocol for Economically
  Sustainable Information Permanence}}.
\newblock \bibinfo{type}{{T}echnical {R}eport}.
  \bibinfo{institution}{arweave.org}.
\newblock


\bibitem[\protect\citeauthoryear{Wolchok and Halderman}{Wolchok and
  Halderman}{2010}]%
        {267312}
\bibfield{author}{\bibinfo{person}{Scott Wolchok} {and}
  \bibinfo{person}{J.~Alex Halderman}.} \bibinfo{year}{2010}\natexlab{}.
\newblock \showarticletitle{Crawling {BitTorrent} {DHTs} for Fun and Profit}.
  In \bibinfo{booktitle}{\emph{4th USENIX Workshop on Offensive Technologies
  (WOOT 10)}}. \bibinfo{publisher}{USENIX Association},
  \bibinfo{address}{Washington, DC}.
\newblock
\urldef\tempurl%
\url{https://www.usenix.org/conference/woot10/crawling-bittorrent-dhts-fun-and-profit}
\showURL{%
\tempurl}


\bibitem[\protect\citeauthoryear{Zhang, Afanasyev, Burke, Jacobson, claffy,
  Crowley, Papadopoulos, Wang, and Zhang}{Zhang et~al\mbox{.}}{2014}]%
        {10.1145/2656877.2656887}
\bibfield{author}{\bibinfo{person}{Lixia Zhang}, \bibinfo{person}{Alexander
  Afanasyev}, \bibinfo{person}{Jeffrey Burke}, \bibinfo{person}{Van Jacobson},
  \bibinfo{person}{kc claffy}, \bibinfo{person}{Patrick Crowley},
  \bibinfo{person}{Christos Papadopoulos}, \bibinfo{person}{Lan Wang}, {and}
  \bibinfo{person}{Beichuan Zhang}.} \bibinfo{year}{2014}\natexlab{}.
\newblock \showarticletitle{Named Data Networking}.
\newblock \bibinfo{journal}{\emph{SIGCOMM Comput. Commun. Rev.}}
  \bibinfo{volume}{44}, \bibinfo{number}{3} (\bibinfo{date}{jul}
  \bibinfo{year}{2014}), \bibinfo{pages}{66–73}.
\newblock
\showISSN{0146-4833}
\urldef\tempurl%
\url{https://doi.org/10.1145/2656877.2656887}
\showDOI{\tempurl}


\bibitem[\protect\citeauthoryear{Zhao, Huang, Stribling, Rhea, Joseph, and
  Kubiatowicz}{Zhao et~al\mbox{.}}{2004}]%
        {zhao2004tapestry}
\bibfield{author}{\bibinfo{person}{Ben~Y Zhao}, \bibinfo{person}{Ling Huang},
  \bibinfo{person}{Jeremy Stribling}, \bibinfo{person}{Sean~C Rhea},
  \bibinfo{person}{Anthony~D Joseph}, {and} \bibinfo{person}{John~D
  Kubiatowicz}.} \bibinfo{year}{2004}\natexlab{}.
\newblock \showarticletitle{{Tapestry: A resilient global-scale overlay for
  service deployment}}.
\newblock \bibinfo{journal}{\emph{IEEE Journal on Selected Areas in
  Communications}} \bibinfo{volume}{22}, \bibinfo{number}{1}
  (\bibinfo{year}{2004}), \bibinfo{pages}{41--53}.
\newblock


\end{thebibliography}

\end{document}